\documentclass[aps,prd,twocolumn,groupedaddress,reprint,noeprint,longbibliography,superscriptaddress,nofootinbib]{revtex4-2}

\usepackage{color}
\usepackage[normalem]{ulem}
\usepackage{cancel}
\usepackage{enumitem}
\usepackage{verbatim}
\usepackage{afterpage}

\newcommand{\myDel}[1]{{\color{red}\ifmmode\text{\sout{$#1$}}\else\sout{#1}\fi}}

\newcommand{\myDelI}[1]{{\color{purple}\ifmmode\text{\sout{$#1$}}\else\sout{#1}\fi}}

\usepackage{amsmath, amssymb}
\usepackage{mathtools}
\renewcommand{\bm}[1]{{\mbox{\boldmath $ #1 $}}}
\allowdisplaybreaks
\usepackage{graphicx}

\usepackage{braket}

\newcommand\abs[1]{\left|#1\right|}
\makeatletter
\renewcommand{\env@cases}[1][@{}l@{\quad}l@{}]{%
	\let\@ifnextchar\new@ifnextchar
	\left\lbrace
	\def\arraystretch{1.2}%
	\array{#1}%
}
\usepackage{tensor}
\newcommand{\idx}{\indices}
\newcommand{\de}{\partial}

\newcommand\numberthis{\addtocounter{equation}{1}\tag{\theequation}}
\usepackage{kbordermatrix}
\renewcommand{\kbldelim}{(}
\renewcommand{\kbrdelim}{)}

\usepackage[utf8]{inputenc}
\usepackage{amsthm}

\usepackage{array}
\usepackage{hhline}
\usepackage{tabularx}
\newcolumntype{Y}{>{\centering\arraybackslash}X}
\newcolumntype{H}{@{}>{\lrbox0}l<{\endlrbox}}
\usepackage{longtable}
\usepackage{makecell}
\usepackage{placeins} 
\def\LT@LR@e{\LTleft\z@   \LTright\z@}%

\usepackage[usestackEOL]{stackengine} 

\usepackage[hidelinks,draft=false]{hyperref}
\usepackage{cleveref} 
\crefname{appch}{Appendix}{Appendices}

\newcounter{magicrownumbers}
\newcommand{\rownumber}{\refstepcounter{magicrownumbers}\arabic{magicrownumbers}}
\newcommand{\refrownumber}[1]{\refstepcounter{magicrownumbers}\label{#1}\arabic{magicrownumbers}}

\usepackage{ifdraft}

\ifdraft{
	\usepackage[colorinlistoftodos]{todonotes}
	
}{
	\usepackage[colorinlistoftodos]{todonotes}
	
}

\AtBeginDocument{%
	\newwrite\bibnotes
	\def\bibnotesext{Notes.bib}
	\immediate\openout\bibnotes=\jobname\bibnotesext
	\immediate\write\bibnotes{@CONTROL{REVTEX41Control}}
	\immediate\write\bibnotes{@CONTROL{%
			apsrev41Control,author="08",editor="1",pages="0",title="0",year="1"}}
	\if@filesw
	\immediate\write\@auxout{\string\citation{apsrev41Control}}%
	\fi
}%

\usepackage{balance}
\usepackage{lastpage}

\AtBeginEnvironment{thebibliography}{\balance}

\begin{document}
	
	
\title{Power-counting renormalizable, ghost-and-tachyon-free Poincar\'e gauge theories}
	
	
	\author{Yun-Cherng~\surname{Lin}}
	\email{ycl54@mrao.cam.ac.uk}
	\affiliation{Astrophysics Group, Cavendish Laboratory, JJ Thomson Avenue,
		Cambridge CB3 0HE, United Kingdom}
	\affiliation{Kavli Institute for Cosmology, Madingley Road, Cambridge CB3 0HA, United Kingdom}
	\author{Michael P.~\surname{Hobson}}
	\email{mph@mrao.cam.ac.uk}
	\affiliation{Astrophysics Group, Cavendish Laboratory, JJ Thomson Avenue,
		Cambridge CB3 0HE, United Kingdom}
	\author{Anthony N.~\surname{Lasenby}}
	\email{a.n.lasenby@mrao.cam.ac.uk}
	\affiliation{Astrophysics Group, Cavendish Laboratory, JJ Thomson Avenue,
		Cambridge CB3 0HE, United Kingdom}
	\affiliation{Kavli Institute for Cosmology, Madingley Road, Cambridge 
		CB3 0HA, United Kingdom}
	
	
	\date{\today}
	
	\begin{abstract}
	We present 48 further examples, in addition to the 10
        identified in \cite{Lin2019a}, of ghost-and-tachyon-free
        critical cases of parity-conserving Poincar\'e gauge
        theories of gravity
        (PGT$^+$) that are also power-counting renormalizable
        (PCR). This is achieved by extending the range of critical
        cases considered. Of the new PCR theories, seven have 2 massless
        degrees of freedom (d.o.f.) in propagating modes and a massive
        $0^-$ or $2^-$ mode, eight have only 2 massless d.o.f., and 33 have only 
        massive mode(s). We also clarify the treatment of
        nonpropagating modes in determining whether a theory is PCR.
	\end{abstract}
	
	
	\maketitle
	

In a recent paper \cite{Lin2019a}, we presented a systematic method
for identifying the ghost-and-tachyon-free critical cases of
parity-preserving gauge theories of gravity and applied it to
parity-preserving Poincar\'e gauge theory (PGT$^+$). The gravitational
free-field Lagrangian for this theory may be written as
\cite{Sezgin1980}
	\begin{align*}
	\frac{\mathcal{L_{\text{G}}}}{\abs{b}}&=
	-\lambda \mathcal{R}
	+\left(r_4+r_5\right) \mathcal{R}^{AB} \mathcal{R}_{AB}\\
	&+\left(r_4-r_5\right) \mathcal{R}^{AB} \mathcal{R}_{BA} 
	+\left(\frac{r_1}{3}+\frac{r_2}{6}\right) \mathcal{R}^{ABCD}
	\mathcal{R}_{ABCD}\\
	&+\left(\frac{2 r_1}{3}-\frac{2 r_2}{3}\right) 
	\mathcal{R}^{ABCD} \mathcal{R}_{ACBD} \\
	&+\left(\frac{r_1}{3}+\frac{r_2}{6}-r_3\right) 
	\mathcal{R}^{ABCD} \mathcal{R}_{CDAB} \\
	&+\left(\frac{\lambda }{4}+\frac{t_1}{3}+\frac{t_2}{12}\right) 
	\mathcal{T}^{ABC} \mathcal{T}_{ABC} \\
	&+\left(-\frac{\lambda }{2}-\frac{t_1}{3}+\frac{t_2}{6}\right) 
	\mathcal{T}^{ABC} \mathcal{T}_{BCA}\\
	&+\left(-\lambda -\frac{t_1}{3}+\frac{2 t_3}{3}\right) 
	\mathcal{T}\text{}_B\text{}^A\text{}^B \mathcal{T}\text{}_C\text{}_A\text{}^C,
	\label{eqn:PGTLagrangian} \numberthis
	\end{align*}
	where the field strengths are
	\begin{align}
	&\mathcal{R}\idx{^{AB}_{\mu\nu}} = 
	2(\de_{[\mu}A\idx{^{AB}_{\nu]}}+A\idx{^A_{E[\mu}}A\idx{^{EB}_{\nu]}}),
	\label{eqn:Rdef} \\
	&\mathcal{T}\idx{^A_{\mu\nu}}=2(\de_{[\mu}b\idx{^A_{\nu]}}
	+A\idx{^A_{E[\mu}}b\idx{^E_{\nu]}}),
	\end{align}
	in which $h\idx{_A^\mu}$ is the translational gauge field,
        $b\idx{^A_\mu}$ is its inverse, such that
        $b\idx{^A_\mu}h\idx{_B^\mu}=\delta^A_B$ and
        $b\idx{^A_\mu}h\idx{_A^\nu}=\delta^\nu_\mu$, and
        $A\idx{^{AB}_\mu} = -A\idx{^{BA}_\mu}$ is the gauge field
        corresponding to Lorentz transformations. Greek indices denote
        the coordinate frame, and Latin capital indices correspond to
        the local Lorentz frame. In our analysis, we linearized the
        gauge fields and decomposed the $h$ field into its symmetric
        and antisymmetric parts $\mathfrak{s}$ and $\mathfrak{a}$,
        respectively, to obtain a quadratic Lagrangian, which we then
        decomposed into
	\begin{align}\mathcal{L}_\mathrm{F} = \sum_{J,P,i,j}a(J^P)_{ij}\hat{\zeta}^\dagger\cdot \hat{P}(J^P)_{ij}\cdot \hat{\zeta}
	\end{align}
	using the spin projection operators (SPOs) $\hat{P}(J^P)_{ij}$
        \cite{Rivers1964,Barnes1965,Aurilia1969}. Please see Sec.
        II of \cite{Lin2019a} for a description of our notation. If
        any of the matrices $a(J^P)$ is singular, then the theory
        possesses gauge invariances. One may fix these gauges by deleting rows and
        columns of the $a$ matrices such that they become
        nonsingular; the elements of the resulting matrices are
        usually denoted by $b_{ij}(J^P)$. The requirement that a
        theory is free from ghosts and tachyons places conditions on
        the $b$ matrices; we traverse all critical cases to determine
        which (if any) satisfy these conditions.

	In this way, in \cite{Lin2019a} we found 450 critical cases
        that are free from ghosts and tachyons, of which we identified
        10 that are also power-counting renormalizable (PCR). The key
        quantity for determining whether a theory is PCR is the propagator
        	\begin{equation}
        	\hat{D} = \sum_{J,P,i,j} b^{-1}_{ij}\hat{P}(J^P)_{ij}.
        	\end{equation}
In particular, if the $b$ matrices contain no elements linking any of
the $A$, $\mathfrak{s}$ and $\mathfrak{a}$ fields , then it is
straightforward to obtain the propagators for these fields separately
from $\hat{D}$.  The original PCR criterion used in \cite{Sezgin1980}
requires the propagator of the $A$-field to decay at least as quickly
as $k^{-2}$ at high energy, and those of the $\mathfrak{s}$ and
$\mathfrak{a}$ fields to fall off at least as $k^{-4}$. By contrast,
the alternative PCR criterion used in \cite{Lin2019a} also permits the
presence of nonpropagating fields (for which the propagator decays no
faster than $\sim k^0$), since these should completely decouple
from the rest of the theory; we will compare these two criteria
further below. Moreover, in \cite{Lin2019a}, we restricted our PCR
considerations to those cases for which the $b$ matrices are diagonal,
such that there are no mixing terms between the $A$, $\mathfrak{s}$
and $\mathfrak{a}$ fields in the gauge-fixed Lagrangian and so the physical
interpretation is straightforward. Indeed, with this restriction, the
high-energy behaviors of the propagators are equivalent to those of
the corresponding diagonal elements in the $b^{-1}$ matrices.
	
	In this paper, we extend our search to include those cases
        for which the $b$ matrices are block diagonal, with each block
        containing only one field. This clearly includes our previous
        study as a special case, but increases considerably the number
        of cases under consideration, while again ensuring that there
        are no mixing terms in the gauge-fixed
        Lagrangian.\footnote{We note that this extension therefore
          does not include Einstein-Cartan theory.}  It is worth
        noting that, even in this more general case, the behavior of
        the propagators at high energy goes as the highest power of
        the corresponding elements in the $b^{-1}$ matrices. Moreover,
        in the PGT$^{+}$ cases we consider, any nondiagonal block of
        a $b$ matrix that does not mix fields is always the only block
        in the matrix, contains only the $A$ field, and has size $2
        \times 2$.  Moreover, these blocks occur only in the $1^-$ or
        $1^+$ sector and have the following form:
		\begin{equation}
		b = \left(\begin{array}{cc}
		r k^2 + (x+4y) & -\sqrt{2}\left(x-2y\right) \\
		-\sqrt{2}\left(x-2y\right) & 2\left(x+y\right)
		\end{array}\right), \label{eqn:nonDiagGeneral}
		\end{equation}
    	where $x$, $y$, and $r$ are real linear combinations of the
        parameters in the Lagrangian. Hence, the element with the
        highest power of $k$ in $b^{-1}$ is always a diagonal element. Note that when $x+y=0$ and $r,x,y\neq 0$, the element with the highest power in
            $b^{-1}$ goes as $k^2$, not $k^{-2}$, even though the highest power in
          $b$ is also $k^2$. This is a similar case to that
            summarized in Eqs. (1.2)--(1.4)
          of \cite{Neville1980}. Since there is no pole in the determinant
          $\det(b)=-18x^2$ in this case, there is no propagating mode
          in this sector. 

Our main result is that, in addition to the 10 PCR cases found in
\cite{Lin2019a}, this new search yields a further 48 cases that are
PCR. For completeness, we list all 58 cases (old and new) in
\Cref{tab:PGTUnitaryAndPCMlMv,tab:PGTUnitaryAndPCMlMv2,tab:PGTUnitaryAndPCMl,tab:PGTUnitaryAndPCMl2,tab:PGTUnitaryAndPCMv,tab:PGTUnitaryAndPCMv2},
in which the old cases are indicated with an asterisk followed by the
old number of the case as given in \cite{Lin2019a}.
Tables~\ref{tab:PGTUnitaryAndPCMlMv2} and
\ref{tab:PGTUnitaryAndPCMlMv} summarize the seven cases with both massless
and massive modes, all of which are new and have 2 massless degrees
of freedom (d.o.f.) in propagating modes and a massive $0^-$ or $2^-$
mode. Tables~\ref{tab:PGTUnitaryAndPCMl2} and
\ref{tab:PGTUnitaryAndPCMl} summarize the 12 cases with only massless
modes, of which eight are new and contain only 2 massless d.o.f. Finally,
Tables~\ref{tab:PGTUnitaryAndPCMv2} and \ref{tab:PGTUnitaryAndPCMv}
summarize the 39 cases with only massive modes, of which 33 are new
. For each set of tables, the first
lists the various conditions for each critical case, and the second
lists the ``particle content'' in terms of the diagonal elements in the
$b^{-1}$ matrix of each spin-parity sector in the sequence
$\{0^-,0^+,1^-,1^+,2^-,2^+\}$.

Since we adopt the PCR criterion in \cite{Lin2019a}, which differs
from the original criterion used in \cite{Sezgin1980} by allowing the
presence of nonpropagating fields, it is worth discussing further the
status of such fields in the determination of whether a theory is
PCR. We begin by noting that an important consequence of allowing the
existence of nonpropagating fields is that whether some critical
cases obey our PCR criterion may depend on the choice of gauge
fixing. For example, in the spin-parity sector $0^+$ in
Case~\ref{row:Case08}, the $a$ matrix is
\begin{equation}
	a(0^+)=
	\bordermatrix{
	~&A& \mathfrak{s}&\mathfrak{s} \cr
	~&2 t_3 & -2 i \sqrt{2} k t_3 & 0 \cr
	~&2 i \sqrt{2} k t_3 & 4 k^2 t_3 & 0 \cr
	~&0 & 0 & 0 \cr
	}, \label{eqn:case3a0p}
\end{equation}
which is singular, indicating the presence of gauge
invariances. One may render this matrix nonsingular by deleting rows
and columns in two different ways, corresponding to two different
gauge fixings, which in this case correspond simply to keeping either the
first or the second column and row.  If one chooses to keep only the
second row and column, then this sector contains only an
$\mathfrak{s}$ field, with a propagator that goes as $\sim k^{-2}$ at
high energy, which thus violates both our alternative PCR criterion
and the original one. Conversely, if one chooses to retain only the
first column and row, then the $0^+$ spin-parity sector contains only
a nonpropagating $A$ field, which we contend is harmless and thus
satisfies our alternative PCR criterion, while violating the original
one. The conclusions regarding PCR are therefore gauge dependent.

Overall, we take the view that a theory is PCR if one can find a gauge
in which it satisfies our PCR criterion, irrespective of the existence
of other gauge choices in which the PCR criterion is violated. The
rationale for this view is that a theory should describe the same
physics independently of which gauge one adopts. Thus, if one uses a
particular gauge to make a physical prediction, then one should, in
principle, be able to draw the same physical conclusion in any other
gauge, although most often not in such a transparent manner. 

We therefore consider the $0^+$ sector of Case~\ref{row:Case08} to satisfy our PCR
criterion, whereas it violates the original one in \cite{Sezgin1980}. Moreover, although
the total propagator for a field is the sum of the propagators across
all sectors, it cannot satisfy either PCR condition if that same
condition is violated by the propagator in any sector individually.
This occurs since the high-energy asymptotic behavior is determined by
the term(s) with the highest power, unless they cancel out, but the
SPO decomposition guarantees that such cancellations cannot happen if
$k^2\neq0$, which is the case we are considering here. Thus, Case~\ref{row:Case08} as
a whole violates the original PCR criterion in \cite{Sezgin1980}
because of the nature of the $0^+$ sector, whereas one finds that it
satisfies our alternative PCR criterion, and is hence listed in
Tables~\ref{tab:PGTUnitaryAndPCMl2} and \ref{tab:PGTUnitaryAndPCMl}.

We now explain why this does not, in fact, lead to a contradiction. If
one chooses to keep only the first column and row in
\eqref{eqn:case3a0p}, the resulting $b^{-1}$ matrix is clearly
\begin{equation}
	b^{-1}(0^+) = \left(\frac{1}{2t_3}\right),
\label{eq:binv0+}
\end{equation}
so the field in this sector is not propagating, and the corresponding
propagator is $\sim k^0$ at high energy. 
The key point, however, is that there is no dynamical term in the
Lagrangian for the field corresponding to (\ref{eq:binv0+}). Thus, one
can integrate out this nonpropagating field in the path integral, which is equivalent to
substituting for it in the Lagrangian using its classical equation of
motion obtained by varying the nonpropagating field. This is most transparently achieved by first introducing
polarization basis vectors to decompose the fields and the SPOs, as
discussed in \Cref{sec:polarizationBasis}.  One then expands the
fields in terms of these basis vectors,
	\begin{equation}
	\ket{A} = \sum_{J,P,i,m} P \bar{A}_{i,J^P,m}\ket{i,J^P,m},
	\end{equation}
	from which one obtains the relation
	\begin{equation}
	\hat{P}_{ji}(J^P) \ket{A} = \bar{A}_{i,J^P,m}\ket{j,J^P,m}. \label{eqn:AbasisExpan}
	\end{equation}
	The Lagrangian corresponding to the $0^+$ sector then becomes
	\begin{equation}
	\mathcal{L}(0^+) = t_3 \bar{A}_{1,0^+,0}^2,
	\end{equation}
	and the equation of motion is simply $\bar{A}_{1,0^+,0}=0$, so
        one can simply ignore this sector. One might alternatively use
        the Lagrangian containing the source current here, so that the
        equation of motion becomes $2t_3
        \bar{A}_{1,0^+,0}=\bar{j}_{1,0^+,0}$, where
        $\bar{j}_{1,0^+,0}$ is appropriate expansion of the source
        current in the polarization. Since we are considering only
        free-field theories, however, the source currents can
        themselves be due only to the gauge fields and thus at least
        quadratic. Hence, these source currents can only affect the
        fields to the next order, so we can neglect them in the
        linearized Lagrangian.

The $1^-$ sector of Case~\ref{row:Case08} can also contain nonpropagating
fields. The $a$ matrix for this sector is
	\begin{align*}
	&a(1^-)= \\
	&2\bordermatrix{
		~& A &A&\mathfrak{s}&\mathfrak{a} \cr
		~&3 k^2 \left(r_1+r_5\right)+2 t_3 &  \sqrt{2} t_3 & - i \sqrt{2} k t_3 &  i \sqrt{2} k t_3 \cr
		~&  \sqrt{2} t_3 &  t_3 & - i k t_3 &  i k t_3 \cr
		~& i \sqrt{2} k t_3 &  i k t_3 &  k^2 t_3 & - k^2 t_3 \cr
		~&- i \sqrt{2} k t_3 & - i k t_3 & - k^2 t_3 & k^2 t_3 \cr
	}, \\ \label{eqn:case3a1n}
	\end{align*}
	which is singular as a result of gauge
        invariances. One may render the matrix nonsingular and
        thereby fix the gauge by, for example, choosing the first two
        rows and columns to form the corresponding $b$ matrix, in
        which case the sector contains a propagating $A$ particle and
        a nonpropagating $A$ particle with some mixing term. The
        resulting determinant is
	\begin{equation}
	\det[b(1^-)] = \frac{4}{3} \left(r_1+r_5\right) t_3 k^2,
	\end{equation}
	so there can only be massless modes in this sector. Using the
        expansion \eqref{eqn:AbasisExpan} to reconstruct the
        Lagrangian corresponding to the $1^-$ sector, one obtains
	\begin{align*}
	\mathcal{L}(1^-) =& -\sum_{m=-1}^{1} \left\{\bar{A}_{1,1^-,m}[-3(r_1+r_5)\de^2 +2t_3]\bar{A}_{1,1^-,m} \right.\\
	&\left.+ 2\sqrt{2}t_3 \bar{A}_{1,1^-,m}\bar{A}_{2,1^-,m} + t_3 \bar{A}^2_{2,1^-,m}\right\}. \numberthis
	\end{align*}
	Hence, it is clear that there is a propagating
        $\bar{A}_{1,1^-,m}$ field that is mixed with a
        $\bar{A}_{2,1^-,m}$ field without a dynamical term. One can
        thus integrate out the latter field using its classical
        equation of motion,
	\begin{equation}
	\bar{A}_{2,1^-,m} = -\sqrt{2}\bar{A}_{1,1^-,m},
	\end{equation}
	and the Lagrangian becomes
	\begin{align*}
	\mathcal{L}(1^-) =& -\sum_{m=-1}^{1} \left\{\bar{A}_{1,1^-,m}[-3(r_1+r_5)\de^2 ]\bar{A}_{1,1^-,m}\right\}. \numberthis
	\end{align*}
	This is consistent with there being no massive mode in this
        sector.  Furthermore, one finds that the effect of integrating
        out the nonpropagating fields in the $0^+$ and $1^-$ sectors
        in Case~\ref{row:Case08} is the same as setting $t_3$ to zero,
        and all the $b$ matrices become exactly the same as those of
        Case~\ref{row:Case09}.  Hence, at least in the free-field case
        we are considering, in which the gauge fields do not couple to
        external matter fields, Case~\ref{row:Case08}
        and~\ref{row:Case09} are actually describing the same
        theory. Moreover, since Case~\ref{row:Case09} may be shown to
        satisfy Sezgin's original PCR criterion in \cite{Sezgin1980},
        there is thus no contradiction in Case~\ref{row:Case08}
        satisfying our alternative PCR criterion. Indeed, the
        alternative criterion allows us to identify
        Case~\ref{row:Case08} as PCR, which would be missed using the
        original PCR criterion.
        
	For all Cases~\ref{row:Case01}--\ref{row:Case58}, one may
        similarly check whether, after integrating out the
        nonpropagating fields, the remaining fields are consistent
        with the particle contents that their determinants of
        $b$ matrices indicate. Because all the $b$ matrices containing
        nonpropagating terms in these cases are in the form of
        \eqref{eqn:nonDiagGeneral}, one can perform this check by
        examining only all the ``special cases'' of the form
        \eqref{eqn:nonDiagGeneral} (including the critical cases and
        those with the parameters making any of the elements zero). We
        find that all of them are consistent.

Moreover, as one might expect, one may show that similar equivalences
as Case~\ref{row:Case08} and~\ref{row:Case09} exist between other
cases. For example, one may further demonstrate in the manner outlined above
that: Case~\ref{row:Case02} is equivalent to Case~\ref{row:Case01};
Cases~\ref{row:Case12},~\ref{row:Case14}, and~\ref{row:Case15} are
equivalent to Case~\ref{row:Case10}; Case~\ref{row:Case16} is
equivalent to Case~\ref{row:Case11}; Case~\ref{row:Case25} is
equivalent to Case~\ref{row:Case26}; Case~\ref{row:Case29} is
equivalent to Case~\ref{row:Case30}; Case~\ref{row:Case37} is
equivalent to Case~\ref{row:Case35}; and Case~\ref{row:Case41} is
equivalent to Case~\ref{row:Case27}. Unfortunately, it is not so
straightforward to establish the equivalences among the other
cases. For the critical cases we do not list in this paper, we
anticipate that there will similarly be some groups of equivalent
cases in the above sense, provided they do not couple to external
matter fields, so that one may simplify the ``tree'' of critical
cases. We leave this analysis for future work.  Nonetheless, we do
find that after integrating out all the nonpropagating fields in
Cases~\ref{row:Case01}--\ref{row:Case58}, all the resulting theories
satisfy the original PCR condition. Hence, allowing for
nonpropagating fields does not violate this criterion in practice.

In conclusion, we have found 48 further critical cases of PGT$^+$ that
are both PCR and free of ghosts and tachyons. This is achieved by
extending the range of critical cases considered beyond those
investigated in \cite{Lin2019a}, which previously identified 10 such
theories. In future work, we plan to investigate all these theories
further, but especially those that possess massless propagating
particles, by considering their phenomenology in the context both of
cosmological and compact object solutions.  Note that while a theory
may pass our PCR criterion, this is no guarantee that the theory is
renormalizable, and this would take independent investigation and the
inclusion of interactions.  Indeed, it is shown in
  \cite{Yo1999a,Yo2001} that linearizing a theory can change its
  structure qualitatively, so that the degrees of freedom and gauge
  invariances may differ.  One must therefore perform a full
  nonlinear analysis to determine whether this is the case for the
  theories considered here. We have also clarified the role played by
nonpropagating modes in determining whether a theory is PCR. We
illustrate this issue further in Appendix~\ref{sec:AppProca}, where we
demonstrate the methods used in this paper in the more familiar and
much simpler cases of the Proca and Stueckelberg theories for a
massive spin-1 particle.

	\begin{acknowledgments}
		We thank Robert Lasenby for helpful discussions.
		Y.-C. Lin acknowledges support from the Ministry of
		Education of Taiwan and the Cambridge Commonwealth,
        European \& International Trust via a Taiwan Cambridge
        Scholarship.
	\end{acknowledgments}
	\medskip
	
	\appendix
	\section{\uppercase{Polarization basis vectors} \label{sec:polarizationBasis}}
	Assuming $k^A=(k^0,0,0,k^3)$, we define the polarization basis vectors for four-vectors as
	\begin{align*}
		&\epsilon^A_{(1^-,1)}=\frac{1}{\sqrt{{2}}}\left( \begin{array}{c} 0\\ 1\\ i\\ 0\\\end{array} \right), \;\;\;\epsilon^A_{(1^-,-1)}=\frac{1}{\sqrt{{2}}}\left( \begin{array}{c} 0\\ -1\\ i\\ 0\\\end{array} \right), \\ &\epsilon^A_{(1^-,0)}=\frac{1}{k}\left( \begin{array}{c} k^3\\ 0\\ 0\\ k^0\\\end{array} \right), \;\;\; \epsilon^A_{(0^+,0)}=\frac{1}{k}\left( \begin{array}{c} k^0\\ 0\\ 0\\ k^3\\\end{array} \right). \numberthis
	\end{align*}
	The basis vectors satisfy the orthonormal and completeness conditions,
	\begin{align}
	\epsilon^{*A}_{(J_1^{P_1},m_1)}\epsilon_{A,(J_2^{P_2},m_2)} &= P_1\delta_{J_1,J_2}\delta_{P_1,P_2}\delta_{m_1,m_2}, \label{eqn:basisOrtho1} \\
	\sum_{J,P,m} P\,\epsilon^{A}_{(J^P,m)}\epsilon^*_{B,(J^P,m)}  &= \delta^{A}_{B}. \label{eqn:basisComplete1}
	\end{align}
	
	For the higher rank tensors, we can apply the addition rules
        for angular momentum. For example, a $(2,0)$-tensor $f^{AB}$ can be decomposed as 
	\begin{align*}
	f^{AB} &\in (\mathbf{0}^+\oplus \mathbf{1}^-)\!\otimes\!(\mathbf{0}^+\oplus \mathbf{1}^-)\\
	&=\!(\mathbf{0}^+\otimes \mathbf{0}^+)\!\oplus\! (\mathbf{0}^+\otimes \mathbf{1}^-)\!\oplus\! (\mathbf{1}^-\otimes \mathbf{0}^+)\!\oplus\! (\mathbf{1}^-\otimes \mathbf{1}^-)\\
	&= \mathbf{0}^+ \oplus \mathbf{1}^- \oplus \mathbf{1}^- \oplus (\mathbf{0}^+\oplus \mathbf{1}^+\oplus \mathbf{2}^+). \numberthis\label{20decomp}
	\end{align*}
	The polarization basis is obtained using
        Clebsch-Gordan coefficients\footnote{We adopt the notation of
          the Particle Data Group, which can be found at
          \url{http://pdg.lbl.gov/2008/reviews/clebrpp.pdf}.}. For
        example, some basis elements
        $\epsilon^{AB}_{(J_1^{P_1},J_2^{P_2},J'^{P'},m_{J'})}$
        for $J_1^{P_1}\otimes J_2^{P_2}$ are
	\begin{align*}
	&\epsilon^{AB}_{(1^-,1^-,2^+,+2)}= \epsilon^{A}_{(1^-,1)}\otimes\epsilon^{B}_{(1^-,1)}, \\
	&\epsilon^{AB}_{(1^-,1^-,2^+,+1)}=
          \!\frac{1}{\sqrt{2}}\!\left(
          \epsilon^{A}_{(1^-,1)}\otimes\epsilon^{B}_{(1^-,0)} +
          \epsilon^{A}_{(1^-,0)}\otimes\epsilon^{B}_{(1^-,1)}
          \right).
\numberthis
	\end{align*} 
Moreover, one can decompose any $(2,0)$ tensor into
$f^{AB}=\mathfrak{s}^{AB}+\mathfrak{a}^{AB}$, where $\mathfrak{s}$ is symmetric and $\mathfrak{a}$ is
antisymmetric. One observes from the Clebsch-Gordan coefficients
table that the $\mathbf{2^+}$ and $\mathbf{0^+}$ sectors are symmetric
in $A$ and $B$, whereas the $\mathbf{1^+}$ sector is
antisymmetric. One may thus make a linear combination of the two
$\mathbf{1^-}$ sectors to obtain a symmetric sector and an antisymmetric
sector,
	\begin{align}
	\epsilon^{AB}_{(\texttt{sym},1^-,m)} &\equiv \frac{1}{\sqrt{2}}\left( \epsilon^{AB}_{(0^+,1^-,1^-,m)} + \epsilon^{AB}_{(1^-,0^+,1^-,m)} \right) \\
	\epsilon^{AB}_{(\texttt{ant},1^-,m)} &\equiv \frac{1}{\sqrt{2}}\left( \epsilon^{AB}_{(0^+,1^-,1^-,m)} - \epsilon^{AB}_{(1^-,0^+,1^-,m)} \right).
	\end{align}
	Hence, we can conclude that the symmetric part of
        (\ref{20decomp}) is $\mathbf{2^+} \oplus \mathbf{1^-} \oplus
        \mathbf{0^+} \oplus \mathbf{0^+}$, which has $5+3+1+1=10$
        degrees of freedom, and the antisymmetric part is
        $\mathbf{1^+} \oplus \mathbf{1^-}$, which has $3+3=6$ degrees
        of freedom, all as expected.
	
	One can similarly decompose the $A^{ABC}$ fields, which are antisymmetric on $A$ and $B$, into 
	\begin{align*}
	A^{ABC} &\in (\mathbf{1}^+\oplus \mathbf{1}^-)\otimes (\mathbf{0}^+\oplus \mathbf{1}^-)\\
	&= \mathbf{1}^+ \oplus (\mathbf{0}^-\oplus\mathbf{1}^-\oplus\mathbf{2}^-)\oplus\mathbf{1}^-\oplus(\mathbf{0}^+\oplus\mathbf{1}^+\oplus\mathbf{2}^+). \\
	&= \mathbf{0}^-\oplus\mathbf{0}^+\oplus 2(\mathbf{1}^-)\oplus 2(\mathbf{1}^+)\oplus \mathbf{2}^-\oplus \mathbf{2}^+,
	\end{align*}
	for which the basis is straightforwardly constructed following
        an analogous approach to that illustrated above.
	
	The bases for higher rank tensors satisfy similar
        orthonormality and completeness conditions to
        \eqref{eqn:basisOrtho1} and \eqref{eqn:basisComplete1}:
	\begin{align}
	&\epsilon^{*\alpha}_{(i_1,J_1^{P_1},m_1)}\epsilon_{\alpha,(i_2,J_2^{P_2},m_2)} = P_1\delta_{i_1,i_2}\delta_{J_1,J_2}\delta_{P_1,P_2}\delta_{m_1,m_2} \label{eqn:basisOrtho} \\
	&\sum_{i,j,P,m}\left(P\,\epsilon^{\alpha}_{(i,J^P,m)}\epsilon^*_{\beta,(i,J^P,m)} \right) = \mathbb{I}^{\alpha}_{\beta}, \label{eqn:basisComplete}
	\end{align}
	where $i$ is the label of the basis in the spin sector $J^P$,
        as there might be more than one basis in a sector. The
        $\alpha$ and $\beta$ indices are shorthand for some generic
        indices, such as $\alpha=A_1A_2...A_n$.
	
	We can write the basis vectors together with its corresponding column vector $\mathbf{e}_{a}$ indicating the field (see (10) in \cite{Lin2019a}) in bra-ket notation $\ket{i,J^P,m}$, and the SPOs in \cite{Lin2019a} are related with those polarization basis vectors by
	\begin{equation}
	\hat{P}_{ij}(J^P) = \sum_{m}\ket{i,J^P,m}\bra{j,J^P,m}. 
	\end{equation}
	Note that the bras and kets here do not denote a quantum
        state, but are used merely to denote the field decomposition
        in a straightforward manner. We are taking inspiration from \cite{Dicus2005, Buoninfante2016} in this section.
	
	\section{\uppercase{Proca and Stueckelberg theories} \label{sec:AppProca}}
	
	In this appendix, we illustrate the methods used in this
        paper in the context of the more familiar and much simpler
        Proca and Stueckelberg theories.

	Proca theory contains a massive vector field $B_\mu$ and has
        the free-field Lagrangian,
	\begin{equation}
	\mathcal{L}_{\text{Pr}} =-\tfrac{1}{4} \left(\de_\mu B_\nu - \de_\nu B_\mu \right) \left(\de^\mu B^\nu - \de^\nu B^\mu \right) + \tfrac{1}{2} m^2 B_\mu B^\mu,
	\end{equation}
	with $m>0$, which has no gauge freedoms. The corresponding SPOs are
	\begin{equation}
	\mathsf{P}(0^+) = \bordermatrix{
		~ & B_{\mu} \cr
		B^*_{\rho} & \Omega_{\mu\rho} 
	}\;,\qquad
	\mathsf{P}(1^-) = \bordermatrix{
		~ & B_{\mu} \cr
		B^*_{\rho} & \Theta_{\mu\rho} 
	},
	\end{equation}
	where $\Omega^{\mu\rho}=k^{\mu}
	k^{\rho}/k^2$, and $\Theta^{\mu\rho}=\eta^{\mu\rho}-k^{\mu} k^{\rho}/k^2$. The $a$ matrices of the theory are
	\begin{equation}
	a(0^+) = \bordermatrix{
		~ & B_{\mu} \cr
		B^*_{\mu} & m^2 
	}\;,\;\;
	a(1^-) = \bordermatrix{
		~ & B_{\mu} \cr
		B^*_{\mu} & -k^2+m^2 
	},
	\end{equation}
	which are identical to the $b$ matrices because there are no
        gauge invariances and source constraints. Therefore, the $0^+$
        sector is nonpropagating and the $1^-$ sector corresponds to
        a $k^{-2}$ propagator. Thus, Proca theory satisfies the
        alternative PCR condition in \cite{Lin2019a}, and hence we
        classify it as PCR.

Conversely, Proca theory clearly violates Sezgin's original PCR condition
in \cite{Sezgin1980}. Indeed, Proca theory is generally considered to
be non-PCR in the literature, because the propagator is
	\begin{equation}
	D(k)_{\mu\nu} = \frac{\eta_{\mu\nu} - \frac{k_\mu k_\nu}{m^2}}{k^2 - m^2},
	\end{equation} 
	so some components of it become $\sim k^0$ when $k^2 \rightarrow
        \infty$ and the offending term $k_\mu k_\nu$ cannot be eliminated by
        the renormalization procedure \cite{Ruegg2003}.  Using the
        polarization basis method mentioned in the main text, however,
        we can integrate out the nonpropagating $0^+$ part. The free
        Lagrangian then becomes $\mathcal{L}_{\text{Pr}}$ with the
        condition $\de^\mu B_\mu=0$, and the resulting propagator goes
        as $k^{-2}$, so the theory is PCR.

One may gain some insight into this apparent contradiction by noting
that Proca theory may be considered as a gauge-fixed version of a
gauge theory, namely the Stueckelberg theory, for which the Lagrangian
is \cite{Stueckelberg1938,Stueckelberg1938a,VanHees2003}
	\begin{align*}
	\mathcal{L}_{\text{St}} =& -\tfrac{1}{4} \left(\de_\mu B_\nu - \de_\nu B_\mu \right) \left(\de^\mu B^\nu - \de^\nu B^\mu \right) + \tfrac{1}{2} m^2 B_\mu B^\mu \\
	&+ \tfrac{1}{2}\de_\mu \phi \de^\mu \phi + m \phi \de_\mu B^\mu \label{eqn:LagStueck} \numberthis
	\end{align*}
	and which possesses the gauge invariance,
	\begin{equation}
	B'_\mu = B_\mu + \de_\mu \Lambda, \qquad \phi' = \phi + m \Lambda.
	\end{equation}
	The nonzero $a$ matrices are
	\begin{align}
	&a(0^+) = \bordermatrix{
		~ & \phi & B_{\mu} \cr
		\phi^* & k^2 & -i k m \cr
		B^*_{\mu} & i k m & m^2 
	},\\
	&a(1^-) = \bordermatrix{
		~ & B_{\mu} \cr
		B^*_{\mu} & -k^2+m^2 
	},
	\end{align}
	and the corresponding SPOs are
	\begin{equation}
	\mathsf{P}(0^+) = \bordermatrix{
		~ & \phi & B_{\mu} \cr
		\phi^* & 1 & \tilde{k}_\mu \cr
		B^*_{\rho} & \tilde{k}_\rho & \Omega_{\mu\rho} 
	}\;,\qquad
	\mathsf{P}(1^-) = \bordermatrix{
		~ & B_{\mu} \cr
		B^*_{\rho} & \Theta_{\mu\rho} 
	},
	\end{equation}
	where $\tilde{k}\text{}_\mu=k\text{}_\mu/\sqrt{k^2}$.  As
        might be expected, the matrix $a(0^+)$ is singular, with rank
        one, and so we can choose to keep either the $\phi$ column/row
        or the $B$ column/row. If we choose to keep $B$, then one
        recovers Proca's theory. If we instead choose to keep $\phi$,
        then the $b^{-1}$ matrices all go as $\sim k^{-2}$ in the
        high-energy limit and the theory thus satisfies the original
        PCR condition. Hence, Stueckelberg theory is PCR, and so Proca
        theory must also be PCR, since the two theories are physically
        equivalent. Thus, our alternative PCR criterion succeeds in
        identifying Proca theory as being PCR, whereas the theory
        violates the original PCR criterion.

	\bgroup \def\arraystretch{1.5}
	\setlength\tabcolsep{0.2cm}
	\begin{longtable*}[e]{@{\extracolsep{\fill}}rlll}
		\caption{Parameter conditions for the PC renormalizable
			critical cases that are ghost and tachyon free and
			have both massless and massive propagating
			modes. The parameters listed in ``Additional
			conditions'' must be nonzero to prevent the
			theory becoming a different critical case.}
		\label{tab:PGTUnitaryAndPCMlMv2} \\*
			\toprule
			\#&Critical condition&Additional conditions&No-ghost-and-tachyon condition\\*
			\colrule
			\noalign{\vspace{3pt}}%
			\endfirsthead
			\noalign{\nobreak\vspace{3pt}}%
			\botrule
			\endlastfoot
			
			\refrownumber{row:Case01} & \makecell[cl]{$r_1,\frac{r_3}{2}-r_4,t_1,t_3,\lambda =0$} &\makecell[cl]{$r_2,r_3,2 r_3+r_5,r_3+2 r_5,t_2$} & $t_2>0, r_2<0, r_3 \left(2 r_3+r_5\right) \left(r_3+2 r_5\right)<0$  \\
			\refrownumber{row:Case02} & \makecell[cl]{$r_1,\frac{r_3}{2}-r_4,t_1,\lambda =0$} &\makecell[cl]{$r_2,r_1-r_3,2 r_3+r_5,r_1+r_3+2 r_5,t_2,t_3$} & $t_2>0, r_2<0, r_3 \left(2 r_3+r_5\right) \left(r_3+2 r_5\right)<0$  \\
			\refrownumber{row:Case03} & \makecell[cl]{$r_1,r_3,r_4,t_1+t_2,t_3,\lambda =0$} &\makecell[cl]{$r_2,r_1+r_5,2 r_1+r_5,t_1,t_2$} & $r_2<0, r_5<0, t_1<0$  \\
			\refrownumber{row:Case04} & \makecell[cl]{$r_2,r_1-r_3,r_4,t_1+t_2,t_3,\lambda =0$} &\makecell[cl]{$r_1,r_1+r_5,2 r_1+r_5,t_1,t_2$} & $t_1>0, r_1+r_5<0, r_1<0$  \\
			\refrownumber{row:Case05} & \makecell[cl]{$r_2,r_1-r_3,r_4,t_2,t_1+t_3,\lambda =0$} &\makecell[cl]{$r_1,r_1+r_5,2 r_1+r_5,t_1,t_3$} & $r_5>0, 2 r_1+r_5>0, t_1>0, r_1<0$  \\
			\refrownumber{row:Case06} & \makecell[cl]{$r_1,2 r_3-r_4,t_1+t_2,t_3,\lambda =0$} &\makecell[cl]{$r_2,r_1-r_3,r_1-2 r_3-r_5,2 r_3+r_5,t_1,t_2$} & $r_2<0, 2 r_3+r_5<0, t_1<0$  \\
			\refrownumber{row:Case07} & \makecell[cl]{$r_2,2 r_1-2 r_3+r_4,t_1+t_2,t_3,\lambda =0$} &\makecell[cl]{$r_1,r_1-r_3,r_1-2 r_3-r_5,2 r_3+r_5,t_1,t_2$} & $t_1>0, r_1<0, 2 r_3+r_5<r_1$  \\
	\end{longtable*}
	\egroup
	
	
	\setcounter{magicrownumbers}{0}
	
	
	\bgroup \def\arraystretch{1.5}
	\setlength\tabcolsep{0.2cm}
	\begin{longtable*}[e]{@{\extracolsep{\fill}}rHccl}
		\caption{Particle content of the
			PC renormalizable critical cases that are
			ghost and tachyon free and have both massless and
			massive propagating modes.  All of these
			cases have 2 massless d.o.f. in propagating modes,
			and also a massive mode. The column ``$b$
			sectors'' describes the diagonal elements in the
			$b^{-1}$ matrix of each spin-parity sector in the
			sequence $\{0^-,0^+,1^-,1^+,2^-,2^+\}$. Here and in
			\Cref{tab:PGTUnitaryAndPCMl,tab:PGTUnitaryAndPCMv} it is notated as
			$\varphi^n_{v}$ or $\varphi^n_{l}$, where $\varphi$
			is the field, $-n$ is the power of $k$ in the
			element in the $b^{-1}$ matrix when $k$ goes to infinity, $v$ means massive
			pole, and $l$ means massless pole. If $n=\infty$, it
			represents that the diagonal element is zero. If
			$n\leq0$, the field is not propagating. The ``$|$''
			notation denotes the different form of the elements
			of the $b^{-1}$ matrices in different choices of
			gauge fixing, and the ``$\&$'' connects the diagonal
			elements in the same $b^{-1}$ matrix. The
			superscript ``N'' represents that there is nonzero
			off-diagonal term in the $b^{-1}$ matrix.}
		\label{tab:PGTUnitaryAndPCMlMv} \\
			\toprule
			\#&Critical condition&\makecell[cl]{Massless\\ mode d.o.f.}&\makecell[cl]{Massive \\mode}&$b$ sectors\\*
			\colrule
			\noalign{\vspace{3pt}}%
			\endfirsthead
			
			\multicolumn{5}{l}{TABLE~\ref{tab:PGTUnitaryAndPCMlMv} (continued)}
			\rule{0pt}{12pt}\\
			\noalign{\vspace{1.5pt}}
			\colrule\rule{0pt}{12pt}
			\#&Critical condition&\makecell[cl]{Massless\\ mode d.o.f.}&\makecell[cl]{Massive \\mode}&$b$ sectors\\*
			\colrule
			\noalign{\vspace{3pt}}%
			\endhead
			\noalign{\nobreak\vspace{3pt}}%
			\colrule
			\endfoot
			\noalign{\nobreak\vspace{3pt}}%
			\botrule
			\endlastfoot
			
			\rownumber & \makecell[cl]{$r_1,\frac{r_3}{2}-r_4,t_1,t_3,\lambda =0$} &2 & $0^-$ & $\left\{A\text{}_{\text{v}}^{2},\times,A\text{}_{\text{l}}^{2},\left(A\text{}_{\text{l}}^{2}\&A\text{}_{\text{l}}^{0}\right)^\text{N}|\left(A\text{}_{\text{l}}^{2}\&\mathfrak{a}\text{}_{\text{l}}^{2}\right)^\text{N},\times,A\text{}_{\text{l}}^{2}\right\}$ \\
			\rownumber & \makecell[cl]{$r_1,\frac{r_3}{2}-r_4,t_1,\lambda =0$} &2 & $0^-$ & $\left\{A\text{}_{\text{v}}^{2},A\text{}_{\text{}}^{0}|\mathfrak{s}\text{}_{\text{l}}^{2},\left(A\text{}_{\text{l}}^{2}\&A\text{}_{\text{l}}^{0}\right)^\text{N}|\left(A\text{}_{\text{l}}^{2}\&\mathfrak{s}\text{}_{\text{l}}^{2}\right)^\text{N}|\left(A\text{}_{\text{l}}^{2}\&\mathfrak{a}\text{}_{\text{l}}^{2}\right)^\text{N},\left(A\text{}_{\text{l}}^{2}\&A\text{}_{\text{l}}^{0}\right)^\text{N}|\left(A\text{}_{\text{l}}^{2}\&\mathfrak{a}\text{}_{\text{l}}^{2}\right)^\text{N},\times,A\text{}_{\text{l}}^{2}\right\}$ \\
			\rownumber & \makecell[cl]{$r_1,r_3,r_4,$\\$t_1+t_2,t_3,\lambda =0$} &2 & $0^-$ & $\left\{A\text{}_{\text{v}}^{2},\times,\left(A\text{}_{\text{l}}^{2}\&A\text{}_{\text{l}}^{0}\right)^\text{N}|\left(A\text{}_{\text{l}}^{2}\&\mathfrak{s}\text{}_{\text{l}}^{2}\right)^\text{N}|\left(A\text{}_{\text{l}}^{2}\&\mathfrak{a}\text{}_{\text{l}}^{2}\right)^\text{N},\left(A\text{}_{\text{}}^{\infty}\&A\text{}_{\text{}}^{-2}\right)^\text{N}|\left(A\text{}_{\text{}}^{\infty}\&\mathfrak{a}\text{}_{\text{l}}^{0}\right)^\text{N},A\text{}_{\text{}}^{0},A\text{}_{\text{}}^{0}|\mathfrak{s}\text{}_{\text{l}}^{2}\right\}$ \\
			\rownumber & \makecell[cl]{$r_2,r_1-r_3,r_4,$\\$t_1+t_2,t_3,\lambda =0$} &2 & $2^-$ & $\left\{A\text{}_{\text{}}^{0},\times,\left(A\text{}_{\text{l}}^{2}\&A\text{}_{\text{l}}^{0}\right)^\text{N}|\left(A\text{}_{\text{l}}^{2}\&\mathfrak{s}\text{}_{\text{l}}^{2}\right)^\text{N}|\left(A\text{}_{\text{l}}^{2}\&\mathfrak{a}\text{}_{\text{l}}^{2}\right)^\text{N},\left(A\text{}_{\text{}}^{\infty}\&A\text{}_{\text{}}^{-2}\right)^\text{N}|\left(A\text{}_{\text{}}^{\infty}\&\mathfrak{a}\text{}_{\text{l}}^{0}\right)^\text{N},A\text{}_{\text{v}}^{2},A\text{}_{\text{}}^{0}|\mathfrak{s}\text{}_{\text{l}}^{2}\right\}$ \\
			\rownumber & \makecell[cl]{$r_2,r_1-r_3,r_4,$\\$t_2,t_1+t_3,\lambda =0$} &2 & $2^-$ & $\left\{\times,A\text{}_{\text{}}^{0}|\mathfrak{s}\text{}_{\text{l}}^{2},\left(A\text{}_{\text{}}^{\infty}\&A\text{}_{\text{}}^{-2}\right)^\text{N}|\left(A\text{}_{\text{}}^{\infty}\&\mathfrak{s}\text{}_{\text{l}}^{0}\right)^\text{N}|\left(A\text{}_{\text{}}^{\infty}\&\mathfrak{a}\text{}_{\text{l}}^{0}\right)^\text{N},\left(A\text{}_{\text{l}}^{2}\&A\text{}_{\text{l}}^{0}\right)^\text{N}|\left(A\text{}_{\text{l}}^{2}\&\mathfrak{a}\text{}_{\text{l}}^{2}\right)^\text{N},A\text{}_{\text{v}}^{2},A\text{}_{\text{}}^{0}|\mathfrak{s}\text{}_{\text{l}}^{2}\right\}$ \\
			\rownumber & \makecell[cl]{$r_1,2 r_3-r_4,$\\$t_1+t_2,t_3,\lambda =0$} &2 & $0^-$ & $\left\{A\text{}_{\text{v}}^{2},A\text{}_{\text{l}}^{2},\left(A\text{}_{\text{l}}^{2}\&A\text{}_{\text{l}}^{0}\right)^\text{N}|\left(A\text{}_{\text{l}}^{2}\&\mathfrak{s}\text{}_{\text{l}}^{2}\right)^\text{N}|\left(A\text{}_{\text{l}}^{2}\&\mathfrak{a}\text{}_{\text{l}}^{2}\right)^\text{N},\left(A\text{}_{\text{}}^{\infty}\&A\text{}_{\text{}}^{-2}\right)^\text{N}|\left(A\text{}_{\text{}}^{\infty}\&\mathfrak{a}\text{}_{\text{l}}^{0}\right)^\text{N},A\text{}_{\text{}}^{0},A\text{}_{\text{}}^{0}|\mathfrak{s}\text{}_{\text{l}}^{2}\right\}$ \\
			\rownumber & \makecell[cl]{$r_2,2 r_1-2 r_3+r_4,$\\$t_1+t_2,t_3,\lambda =0$} &2 & $2^-$ & $\left\{A\text{}_{\text{}}^{0},A\text{}_{\text{l}}^{2},\left(A\text{}_{\text{l}}^{2}\&A\text{}_{\text{l}}^{0}\right)^\text{N}|\left(A\text{}_{\text{l}}^{2}\&\mathfrak{s}\text{}_{\text{l}}^{2}\right)^\text{N}|\left(A\text{}_{\text{l}}^{2}\&\mathfrak{a}\text{}_{\text{l}}^{2}\right)^\text{N},\left(A\text{}_{\text{}}^{\infty}\&A\text{}_{\text{}}^{-2}\right)^\text{N}|\left(A\text{}_{\text{}}^{\infty}\&\mathfrak{a}\text{}_{\text{l}}^{0}\right)^\text{N},A\text{}_{\text{v}}^{2},A\text{}_{\text{}}^{0}|\mathfrak{s}\text{}_{\text{l}}^{2}\right\}$ \\
	\end{longtable*}
	\egroup
	

	\newcounter{tmp}
	\setcounter{tmp}{\value{magicrownumbers}}
	
	
	\bgroup \def\arraystretch{1.5}
	\begin{longtable*}[e]{@{\extracolsep{\fill}}rlll}
		\caption{Parameter conditions for the PC
			renormalizable critical cases that are ghost and
			tachyon free and have only massless propagating
			modes. The cases found previously in \cite{Lin2019a}
			are indicated with an asterisk followed by its
			original numbering.}
		\label{tab:PGTUnitaryAndPCMl2}\\
			\toprule
			\#&Critical condition&Additional condition&No-ghost-and-tachyon condition\\*
			\colrule
			\noalign{\vspace{3pt}}%
			\endfirsthead
			
			\multicolumn{4}{l}{TABLE~\ref{tab:PGTUnitaryAndPCMl2} (continued)}
			\rule{0pt}{12pt}\\
			\noalign{\vspace{1.5pt}}
			\colrule\rule{0pt}{12pt}
			\#&Critical Condition&Additional Condition&No-ghost-and-tachyon Condition\\*
			\colrule
			\noalign{\vspace{3pt}}%
			\endhead
			\noalign{\nobreak\vspace{3pt}}%
			\colrule
			\endfoot
			\noalign{\nobreak\vspace{3pt}}%
			\botrule
			\endlastfoot
			
			\refrownumber{row:Case08} & \makecell[cl]{$r_2,r_1-r_3,r_4,t_1,t_2,\lambda =0$} &\makecell[cl]{$r_1,r_1+r_5,2 r_1+r_5,t_3$} & $r_1 \left(r_1+r_5\right) \left(2 r_1+r_5\right)<0$  \\
			$^{\ast 1}$\refrownumber{row:Case09} & \makecell[cl]{$r_2,r_1-r_3,r_4,t_1,t_2,t_3,\lambda =0$} &\makecell[cl]{$r_1,r_1+r_5,2 r_1+r_5$} & $r_1 \left(r_1+r_5\right) \left(2 r_1+r_5\right)<0$  \\
			$^{\ast 3}$\refrownumber{row:Case10} & \makecell[cl]{$r_1,r_2,\frac{r_3}{2}-r_4,t_1,t_2,t_3,\lambda =0$} &\makecell[cl]{$r_3,2 r_3+r_5,r_3+2 r_5$} & $r_3 \left(2 r_3+r_5\right) \left(r_3+2 r_5\right)<0$  \\
			$^{\ast 4}$\refrownumber{row:Case11} & \makecell[cl]{$r_1,\frac{r_3}{2}-r_4,t_1,t_2,t_3,\lambda =0$} &\makecell[cl]{$r_2,r_3,2 r_3+r_5,r_3+2 r_5$} & $r_3 \left(2 r_3+r_5\right) \left(r_3+2 r_5\right)<0$  \\
			\refrownumber{row:Case12} & \makecell[cl]{$r_1,r_2,\frac{r_3}{2}-r_4,t_1,t_3,\lambda =0$} &\makecell[cl]{$r_3,2 r_3+r_5,r_3+2 r_5,t_2$} & $r_3 \left(2 r_3+r_5\right) \left(r_3+2 r_5\right)<0$  \\
			$^{\ast 2}$\refrownumber{row:Case13} & \makecell[cl]{$r_2,2 r_1-2 r_3+r_4,t_1,t_2,t_3,\lambda =0$} &\makecell[cl]{$r_1,r_1-r_3,r_1-2 r_3-r_5,2 r_3+r_5$} & $r_1 \left(r_1-2 r_3-r_5\right) \left(2 r_3+r_5\right)>0$  \\
			\refrownumber{row:Case14} & \makecell[cl]{$r_1,r_2,\frac{r_3}{2}-r_4,t_1,t_2,\lambda =0$} &\makecell[cl]{$2 r_3-r_4,2 r_3+r_5,r_4+r_5,t_3$} & $r_3 \left(2 r_3+r_5\right) \left(r_3+2 r_5\right)<0$  \\
			\refrownumber{row:Case15} & \makecell[cl]{$r_1,r_2,\frac{r_3}{2}-r_4,t_1,\lambda =0$} &\makecell[cl]{$r_3,2 r_3+r_5,r_3+2 r_5,t_2,t_3$} & $r_3 \left(2 r_3+r_5\right) \left(r_3+2 r_5\right)<0$  \\
			\refrownumber{row:Case16} & \makecell[cl]{$r_1,\frac{r_3}{2}-r_4,t_1,t_2,\lambda =0$} &\makecell[cl]{$r_2,r_3,2 r_3+r_5,r_3+2 r_5,t_3$} & $r_3 \left(2 r_3+r_5\right) \left(r_3+2 r_5\right)<0$  \\
			\refrownumber{row:Case17} & \makecell[cl]{$r_1,r_2,r_3,r_4,t_1+t_2,t_3,\lambda =0$} &\makecell[cl]{$r_1+r_5,2 r_1+r_5,t_1,t_2$} & $r_5<0, t_1\neq 0$  \\
			\refrownumber{row:Case18} & \makecell[cl]{$r_1,r_2,r_3,r_4,t_2,t_1+t_3,\lambda =0$} &\makecell[cl]{$r_1+r_5,2 r_1+r_5,t_1,t_3$} & $r_5>0, t_1\neq 0$  \\
			\refrownumber{row:Case19} & \makecell[cl]{$r_1,r_2,2 r_3-r_4,t_1+t_2,t_3,\lambda =0$} &\makecell[cl]{$r_1-r_3,r_1-2 r_3-r_5,2 r_3+r_5,t_1,t_2$} & $r_3<-\frac{r_5}{2}, t_1\neq 0$  \\
	\end{longtable*}
	\egroup
	
	
	\setcounter{magicrownumbers}{\value{tmp}}
	
	
	\bgroup \def\arraystretch{1.5}
	\setlength\tabcolsep{0.2cm}
	\begin{longtable*}[e]{@{\extracolsep{\fill}}rHcl}
		\caption{Particle content of the PC renormalizable
			critical cases that are ghost and tachyon free and
			have only massless propagating modes. All
			of these cases have 2 massless d.o.f. of propagating
			mode. The cases found previously in \cite{Lin2019a}
			are indicated with an asterisk followed by its
			original numbering.}
		\label{tab:PGTUnitaryAndPCMl}\\
			\toprule
			\#&Critical Condition&\makecell[cl]{Massless\\ mode d.o.f.}&$b$ sectors\\*
			\colrule
			\noalign{\vspace{3pt}}%
			\endfirsthead
			
			\multicolumn{4}{l}{TABLE~\ref{tab:PGTUnitaryAndPCMl} (continued)}
			\rule{0pt}{12pt}\\
			\noalign{\vspace{1.5pt}}
			\colrule\rule{0pt}{12pt}
			\#&Critical Condition&\makecell[cl]{Massless\\ mode d.o.f.}&$b$ sectors\\*
			\colrule
			\noalign{\vspace{3pt}}%
			\endhead
			\noalign{\nobreak\vspace{3pt}}%
			\colrule
			\endfoot
			\noalign{\nobreak\vspace{3pt}}%
			\botrule
			\endlastfoot
			
			\refrownumber{row:Case3bSec} & \makecell[cl]{$r_2=r_1-r_3=r_4=t_1=t_2=\lambda =0$} & 2 & $\left\{\times,A\text{}_{\text{}}^{0}|\mathfrak{s}\text{}_{\text{l}}^{2},\left(A\text{}_{\text{l}}^{2}\&A\text{}_{\text{l}}^{0}\right)^\text{N}|\left(A\text{}_{\text{l}}^{2}\&\mathfrak{s}\text{}_{\text{l}}^{2}\right)^\text{N}|\left(A\text{}_{\text{l}}^{2}\&\mathfrak{a}\text{}_{\text{l}}^{2}\right)^\text{N},A\text{}_{\text{l}}^{2},A\text{}_{\text{l}}^{2},\times\right\}$ \\
			$^{\ast 1}$\rownumber & \makecell[cl]{$r_2=r_1-r_3=r_4=t_1=t_2=t_3=\lambda =0$} & 2 & $\left\{\times,\times,A\text{}_{\text{l}}^{2},A\text{}_{\text{l}}^{2},A\text{}_{\text{l}}^{2},\times\right\}$ \\
			$^{\ast 3}$\rownumber & \makecell[cl]{$r_1=r_2=\frac{r_3}{2}-r_4=t_1=t_2=t_3=\lambda =0$} & 2 & $\left\{\times,\times,A\text{}_{\text{l}}^{2},A\text{}_{\text{l}}^{2},\times,A\text{}_{\text{l}}^{2}\right\}$ \\
			$^{\ast 4}$\rownumber & \makecell[cl]{$r_1=\frac{r_3}{2}-r_4=t_1=t_2=t_3=\lambda =0$} & 2 & $\left\{A\text{}_{\text{l}}^{2},\times,A\text{}_{\text{l}}^{2},A\text{}_{\text{l}}^{2},\times,A\text{}_{\text{l}}^{2}\right\}$ \\
			\refrownumber{row:Case7bSec} & \makecell[cl]{$r_1=r_2=\frac{r_3}{2}-r_4=t_1=t_3=\lambda =0$} &2 &  $\left\{A\text{}_{\text{}}^{0},\times,A\text{}_{\text{l}}^{2},\left(A\text{}_{\text{l}}^{2}\&A\text{}_{\text{l}}^{0}\right)^\text{N}|\left(A\text{}_{\text{l}}^{2}\&\mathfrak{a}\text{}_{\text{l}}^{2}\right)^\text{N},\times,A\text{}_{\text{l}}^{2}\right\}$ \\
			$^{\ast 2}$\rownumber & \makecell[cl]{$r_2=2 r_1-2 r_3+r_4=t_1=t_2=t_3=\lambda =0$} & 2 & $\left\{\times,A\text{}_{\text{l}}^{2},A\text{}_{\text{l}}^{2},A\text{}_{\text{l}}^{2},A\text{}_{\text{l}}^{2},\times\right\}$ \\
			\rownumber & \makecell[cl]{$r_1=r_2=\frac{r_3}{2}-r_4=t_1=t_2=\lambda =0$} & 2 & $\left\{\times,A\text{}_{\text{}}^{0}|\mathfrak{s}\text{}_{\text{l}}^{2},\left(A\text{}_{\text{l}}^{2}\&A\text{}_{\text{l}}^{0}\right)^\text{N}|\left(A\text{}_{\text{l}}^{2}\&\mathfrak{s}\text{}_{\text{l}}^{2}\right)^\text{N}|\left(A\text{}_{\text{l}}^{2}\&\mathfrak{a}\text{}_{\text{l}}^{2}\right)^\text{N},A\text{}_{\text{l}}^{2},\times,A\text{}_{\text{l}}^{2}\right\}$ \\
			\rownumber & \makecell[cl]{$r_1=r_2=\frac{r_3}{2}-r_4=t_1=\lambda =0$} & 2 & \makecell[cl]{$\left\{A\text{}_{\text{}}^{0},A\text{}_{\text{}}^{0}|\mathfrak{s}\text{}_{\text{l}}^{2},\left(A\text{}_{\text{l}}^{2}\&A\text{}_{\text{l}}^{0}\right)^\text{N}|\left(A\text{}_{\text{l}}^{2}\&\mathfrak{s}\text{}_{\text{l}}^{2}\right)^\text{N}|\left(A\text{}_{\text{l}}^{2}\&\mathfrak{a}\text{}_{\text{l}}^{2}\right)^\text{N},\right.$\\ $\left.\left(A\text{}_{\text{l}}^{2}\&A\text{}_{\text{l}}^{0}\right)^\text{N}|\left(A\text{}_{\text{l}}^{2}\&\mathfrak{a}\text{}_{\text{l}}^{2}\right)^\text{N},\times,A\text{}_{\text{l}}^{2}\right\}$} \\
			\rownumber & \makecell[cl]{$r_1=\frac{r_3}{2}-r_4=t_1=t_2=\lambda =0$} & 2 & $\left\{A\text{}_{\text{l}}^{2},A\text{}_{\text{}}^{0}|\mathfrak{s}\text{}_{\text{l}}^{2},\left(A\text{}_{\text{l}}^{2}\&A\text{}_{\text{l}}^{0}\right)^\text{N}|\left(A\text{}_{\text{l}}^{2}\&\mathfrak{s}\text{}_{\text{l}}^{2}\right)^\text{N}|\left(A\text{}_{\text{l}}^{2}\&\mathfrak{a}\text{}_{\text{l}}^{2}\right)^\text{N},A\text{}_{\text{l}}^{2},\times,A\text{}_{\text{l}}^{2}\right\}$ \\
			\rownumber & \makecell[cl]{$r_1,r_2,r_3,r_4,t_1+t_2,t_3,\lambda =0$} &2 &  $\left\{A\text{}_{\text{}}^{0},\times,\left(A\text{}_{\text{l}}^{2}\&A\text{}_{\text{l}}^{0}\right)^\text{N}|\left(A\text{}_{\text{l}}^{2}\&\mathfrak{s}\text{}_{\text{l}}^{2}\right)^\text{N}|\left(A\text{}_{\text{l}}^{2}\&\mathfrak{a}\text{}_{\text{l}}^{2}\right)^\text{N},\left(A\text{}_{\text{}}^{\infty}\&A\text{}_{\text{}}^{-2}\right)^\text{N}|\left(A\text{}_{\text{}}^{\infty}\&\mathfrak{a}\text{}_{\text{l}}^{0}\right)^\text{N},A\text{}_{\text{}}^{0},A\text{}_{\text{}}^{0}|\mathfrak{s}\text{}_{\text{l}}^{2}\right\}$ \\
			\rownumber & \makecell[cl]{$r_1,r_2,r_3,r_4,t_2,t_1+t_3,\lambda =0$} &2 &  $\left\{\times,A\text{}_{\text{}}^{0}|\mathfrak{s}\text{}_{\text{l}}^{2},\left(A\text{}_{\text{}}^{\infty}\&A\text{}_{\text{}}^{-2}\right)^\text{N}|\left(A\text{}_{\text{}}^{\infty}\&\mathfrak{s}\text{}_{\text{l}}^{0}\right)^\text{N}|\left(A\text{}_{\text{}}^{\infty}\&\mathfrak{a}\text{}_{\text{l}}^{0}\right)^\text{N},\left(A\text{}_{\text{l}}^{2}\&A\text{}_{\text{l}}^{0}\right)^\text{N}|\left(A\text{}_{\text{l}}^{2}\&\mathfrak{a}\text{}_{\text{l}}^{2}\right)^\text{N},A\text{}_{\text{}}^{0},A\text{}_{\text{}}^{0}|\mathfrak{s}\text{}_{\text{l}}^{2}\right\}$ \\
			\rownumber & \makecell[cl]{$r_1,r_2,2 r_3-r_4,t_1+t_2,t_3,\lambda =0$} &2 &  $\left\{A\text{}_{\text{}}^{0},A\text{}_{\text{l}}^{2},\left(A\text{}_{\text{l}}^{2}\&A\text{}_{\text{l}}^{0}\right)^\text{N}|\left(A\text{}_{\text{l}}^{2}\&\mathfrak{s}\text{}_{\text{l}}^{2}\right)^\text{N}|\left(A\text{}_{\text{l}}^{2}\&\mathfrak{a}\text{}_{\text{l}}^{2}\right)^\text{N},\left(A\text{}_{\text{}}^{\infty}\&A\text{}_{\text{}}^{-2}\right)^\text{N}|\left(A\text{}_{\text{}}^{\infty}\&\mathfrak{a}\text{}_{\text{l}}^{0}\right)^\text{N},A\text{}_{\text{}}^{0},A\text{}_{\text{}}^{0}|\mathfrak{s}\text{}_{\text{l}}^{2}\right\}$ \\
	\end{longtable*}
	\egroup
	

	\setcounter{tmp}{\value{magicrownumbers}}
	
	
	\bgroup \def\arraystretch{1.5}
	\setlength\tabcolsep{0.2cm}
	\begin{longtable*}[e]{@{\extracolsep{\fill}}rlll}
		\caption{Parameter conditions for the PC
			renormalizable critical cases that are ghost and
			tachyon free and have only massive propagating
			modes. The cases found previously in \cite{Lin2019a}
			are indicated with an asterisk followed by its
			original numbering.}
		\label{tab:PGTUnitaryAndPCMv2}\\
			\toprule
			\#&Critical condition&Additional conditions&No-ghost-and-tachyon condition\\*
			\colrule
			\noalign{\vspace{3pt}}%
			\endfirsthead
			
			\multicolumn{4}{l}{TABLE~\ref{tab:PGTUnitaryAndPCMv2} (continued)}
			\rule{0pt}{12pt}\\
			\noalign{\vspace{1.5pt}}
			\colrule\rule{0pt}{12pt}
			\#&Critical condition&Additional conditions&No-ghost-and-tachyon condition\\*
			\colrule
			\noalign{\vspace{3pt}}%
			\endhead
			\noalign{\nobreak\vspace{3pt}}%
			\colrule
			\endfoot
			\noalign{\nobreak\vspace{3pt}}%
			\botrule
			\endlastfoot
			
			\refrownumber{row:Case20} & \makecell[cl]{$r_1,r_3,r_4,r_5,\lambda =0$} &\makecell[cl]{$r_2,t_1,t_2,t_1+t_2,t_3,t_1+t_3$} & $t_2>0, r_2<0$  \\
			\refrownumber{row:Case21} & \makecell[cl]{$r_1,r_3,r_4,r_5,t_1+t_2,\lambda =0$} &\makecell[cl]{$r_2,t_1,t_2,t_3,t_1+t_3$} & $r_2<0, t_1<0$  \\
			\refrownumber{row:Case22} & \makecell[cl]{$r_1,r_3,r_4,r_5,t_1+t_3,\lambda =0$} &\makecell[cl]{$r_2,t_1,t_2,t_1+t_2,t_3$} & $t_2>0, r_2<0$  \\
			\refrownumber{row:Case23} & \makecell[cl]{$r_1,r_3,r_4,r_5,t_1+t_2,t_1+t_3,\lambda =0$} &\makecell[cl]{$r_2,t_1,t_2,t_3$} & $r_2<0, t_1<0$  \\
			\refrownumber{row:Case24} & \makecell[cl]{$r_1,r_3,r_4,t_1,\lambda =0$} &\makecell[cl]{$r_2,r_1+r_5,2 r_1+r_5,t_2,t_3$} & $t_2>0, r_2<0$  \\
			$^{\ast 5}$\refrownumber{row:Case25} & \makecell[cl]{$r_1,r_3,r_4,r_5,t_1,\lambda =0$} &\makecell[cl]{$r_2,t_2,t_3$} & $t_2>0, r_2<0$  \\
			$^{\ast 6}$\refrownumber{row:Case26} & \makecell[cl]{$r_1,r_3,r_4,r_5,t_1,t_3,\lambda =0$} &\makecell[cl]{$r_2,t_2$} & $t_2>0, r_2<0$  \\
			\refrownumber{row:Case27} & \makecell[cl]{$r_1,\frac{r_3}{2}-r_4,\frac{r_3}{2}+r_5,t_1,t_3,\lambda =0$} &\makecell[cl]{$r_2,r_3,t_2$} & $t_2>0, r_2<0$  \\
			\refrownumber{row:Case28} & \makecell[cl]{$r_1,r_3,r_4,t_1,t_3,\lambda =0$} &\makecell[cl]{$r_2,r_5,t_2$} & $t_2>0, r_2<0$  \\
			\refrownumber{row:Case29} & \makecell[cl]{$r_1-r_3,r_4,2 r_1+r_5,t_1,\lambda =0$} &\makecell[cl]{$r_1,r_2,r_1+r_5,t_2,t_3$} & $t_2>0, r_2<0$  \\
			$^{\ast 7}$\refrownumber{row:Case30} & \makecell[cl]{$r_1-r_3,r_4,2 r_1+r_5,t_1,t_3,\lambda =0$} &\makecell[cl]{$r_1,r_2,t_2$} & $t_2>0, r_2<0$  \\
			$^{\ast 8}$\refrownumber{row:Case31} & \makecell[cl]{$r_1,2 r_3-r_4,2 r_3+r_5,t_1,t_3,\lambda =0$} &\makecell[cl]{$r_2,r_3,t_2$} & $t_2>0, r_2<0$  \\
			\refrownumber{row:Case32} & \makecell[cl]{$r_1,r_3,r_4,r_5,t_3,\lambda =0$} &\makecell[cl]{$r_2,t_1,t_2,t_1+t_2$} & $t_2>0, r_2<0$  \\
			\refrownumber{row:Case33} & \makecell[cl]{$r_1,r_3,r_4,r_5,t_1+t_2,t_3,\lambda =0$} &\makecell[cl]{$r_2,t_1,t_2$} & $r_2<0, t_1<0$  \\
			\refrownumber{row:Case34} & \makecell[cl]{$r_1,2 r_3-r_4,t_1,t_3,\lambda =0$} &\makecell[cl]{$r_2,r_3,2 r_3+r_5,t_2$} & $t_2>0, r_2<0$  \\
			$^{\ast 9}$\refrownumber{row:Case35} & \makecell[cl]{$r_1,\frac{r_3}{2}-r_4,2 r_3+r_5,t_1,t_3,\lambda =0$} &\makecell[cl]{$r_2,r_3,t_2$} & $t_2>0, r_2<0$  \\
			$^{\ast 10}$\refrownumber{row:Case36} & \makecell[cl]{$2 r_1-2 r_3+r_4,2 r_3+r_5,t_1,t_3,\lambda =0$} &\makecell[cl]{$r_1,r_2,r_1-r_3,t_2$} & $t_2>0, r_2<0$  \\
			\refrownumber{row:Case37} & \makecell[cl]{$r_1,\frac{r_3}{2}-r_4,2 r_3+r_5,t_1,\lambda =0$} &\makecell[cl]{$r_2,2 r_3-r_4,t_2,t_3$} & $t_2>0, r_2<0$  \\
			\refrownumber{row:Case38} & \makecell[cl]{$r_1,2 r_3-r_4,2 r_3+r_5,t_3,\lambda =0$} &\makecell[cl]{$r_2,r_1-r_3,t_1,t_2,t_1+t_2$} & $t_2>0, r_2<0$  \\
			\refrownumber{row:Case39} & \makecell[cl]{$r_1,2 r_3-r_4,2 r_3+r_5,t_1+t_2,t_3,\lambda =0$} &\makecell[cl]{$r_2,r_1-r_3,t_1,t_2$} & $r_2<0, t_1<0$  \\
			\refrownumber{row:Case40} & \makecell[cl]{$r_1,r_4+r_5,t_1,t_3,\lambda =0$} &\makecell[cl]{$r_2,r_3-2 r_4,2 r_3-r_4,t_2$} & $t_2>0, r_2<0$  \\
			\refrownumber{row:Case41} & \makecell[cl]{$r_1,\frac{r_3}{2}-r_4,\frac{r_3}{2}+r_5,t_1,\lambda =0$} &\makecell[cl]{$r_2,2 r_3-r_4,t_2,t_3$} & $t_2>0, r_2<0$  \\
			\refrownumber{row:Case42} & \makecell[cl]{$r_1,r_3,r_4,t_1+t_2,\lambda =0$} &\makecell[cl]{$r_2,r_1+r_5,2 r_1+r_5,t_1,t_2,t_3,$\\$t_1+t_3$} & \makecell[cl]{$t_3>0, r_2<0, r_5<0,$\\$ t_1<0, t_1+t_3<0$}  \\
			\refrownumber{row:Case43} & \makecell[cl]{$r_1,r_3,r_4,t_1+t_3,\lambda =0$} &\makecell[cl]{$r_2,r_1+r_5,2 r_1+r_5,t_1,t_2,$\\$t_1+t_2,t_3$} & \makecell[cl]{$r_5>0, t_2>0, t_1+t_2>0,$\\$ r_2<0, t_1<0$}  \\
			\refrownumber{row:Case44} & \makecell[cl]{$r_2,r_1-r_3,r_4,t_1+t_2,\lambda =0$} &\makecell[cl]{$r_1,r_1+r_5,2 r_1+r_5,t_1,t_2,t_3,$\\$t_1+t_3$} & \makecell[cl]{$t_1>0, r_1<0, r_1+r_5<0,$\\$ t_3 \left(t_1+t_3\right)>0$}  \\
			\refrownumber{row:Case45} & \makecell[cl]{$r_2,r_1-r_3,r_4,t_1+t_3,\lambda =0$} &\makecell[cl]{$r_1,r_1+r_5,2 r_1+r_5,t_1,t_2,$\\$t_1+t_2,t_3$} & \makecell[cl]{$r_5>0, 2 r_1+r_5>0, t_1>0,$\\$ t_1+t_2>0, r_1<0, t_2<0$} \\
			\refrownumber{row:Case46} & \makecell[cl]{$r_1-r_3,r_4,2 r_1+r_5,t_1+t_3,\lambda =0$} &\makecell[cl]{$r_1,r_2,r_1+r_5,t_1,t_2,t_1+t_2,t_3$} & $t_1>0, t_2>0, r_1<0, r_2<0$  \\
			\refrownumber{row:Case47} & \makecell[cl]{$r_1,r_2,r_3,r_4,t_1+t_2,\lambda =0$} &\makecell[cl]{$r_1+r_5,2 r_1+r_5,t_1,t_2,t_3,t_1+t_3$} & $r_5<0, t_1 t_3 \left(t_1+t_3\right)>0$  \\
			\refrownumber{row:Case48} & \makecell[cl]{$r_1,r_2,r_3,r_4,t_1+t_3,\lambda =0$} &\makecell[cl]{$r_1+r_5,2 r_1+r_5,t_1,t_2,t_1+t_2,t_3$} & $r_5>0, t_1 t_2 \left(t_1+t_2\right)<0$  \\
			\refrownumber{row:Case49} & \makecell[cl]{$r_1,r_3,r_4,t_1+t_2,t_1+t_3,\lambda =0$} &\makecell[cl]{$r_2,r_1+r_5,2 r_1+r_5,t_1,t_2,t_3$} & $r_2<0, t_1<0$  \\
			\refrownumber{row:Case50} & \makecell[cl]{$r_2,r_1-r_3,r_4,r_1+r_5,t_1+t_2,\lambda =0$} &\makecell[cl]{$r_1,2 r_1+r_5,t_1,t_2,t_3,t_1+t_3$} & $t_1>0, r_1<0$  \\
			\refrownumber{row:Case51} & \makecell[cl]{$r_2,r_1-r_3,r_4,2 r_1+r_5,t_1+t_3,\lambda =0$} &\makecell[cl]{$r_1,r_1+r_5,t_1,t_2,t_1+t_2,t_3$} & $t_1>0, r_1<0$  \\
			\refrownumber{row:Case52} & \makecell[cl]{$r_2,r_1-r_3,r_4,t_1+t_2,t_1+t_3,\lambda =0$} &\makecell[cl]{$r_1,r_1+r_5,2 r_1+r_5,t_1,t_2,t_3$} & $t_1>0, r_1<0$  \\
			\refrownumber{row:Case53} & \makecell[cl]{$r_2,r_1-r_3,r_4,r_1+r_5,t_1+t_2,t_1+t_3,\lambda =0$} &\makecell[cl]{$r_1,2 r_1+r_5,t_1,t_2,t_3$} & $t_1>0, r_1<0$  \\
			\refrownumber{row:Case54} & \makecell[cl]{$r_2,r_1-r_3,r_4,2 r_1+r_5,t_1+t_2,t_1+t_3,\lambda =0$} &\makecell[cl]{$r_1,r_1+r_5,t_1,t_2,t_3$} & $t_1>0, r_1<0$  \\
			\refrownumber{row:Case55} & \makecell[cl]{$r_2,r_1-r_3,r_4,r_1+r_5,t_1+t_2,t_3,\lambda =0$} &\makecell[cl]{$r_1,t_1,t_2$} & $t_1>0, r_1<0$  \\
			\refrownumber{row:Case56} & \makecell[cl]{$r_2,r_1-r_3,r_4,2 r_1+r_5,t_2,t_1+t_3,\lambda =0$} &\makecell[cl]{$r_1,t_1,t_3$} & $t_1>0, r_1<0$  \\
			\refrownumber{row:Case57} & \makecell[cl]{$r_1-r_3,r_4,2 r_1+r_5,t_2,t_1+t_3,\lambda =0$} &\makecell[cl]{$r_1,r_2,t_1,t_3$} & $t_1>0, r_1<0$  \\
			\refrownumber{row:Case58} & \makecell[cl]{$r_2,2 r_1-2 r_3+r_4,r_1-2 r_3-r_5,$\\$t_1+t_2,t_3,\lambda =0$} &\makecell[cl]{$r_1,r_1-r_3,t_1,t_2$} & $t_1>0, r_1<0$  \\
	\end{longtable*}
	\egroup
	
	
	\setcounter{magicrownumbers}{\value{tmp}}
	
	
	\bgroup \def\arraystretch{1.5}
	\setlength\tabcolsep{0.11cm}
	\begin{longtable*}[e]{@{\extracolsep{\fill}}rHcl}
		\caption{Particle content of the PC renormalizable
			critical cases that are ghost and tachyon free and
			have only massive propagating modes. The cases found previously in \cite{Lin2019a}
			are indicated with an asterisk followed by its
			original numbering. Note that there are typos of the $b$ sectors of Cases~\ref{row:Case30} and~\ref{row:Case31} (old numbers 7 and 8) in \cite{Lin2019a}.}
		\label{tab:PGTUnitaryAndPCMv}\\
			\toprule
			\#&Critical Condition&\makecell[cl]{Massive \\mode}&$b$ sectors\\*
			\colrule
			\noalign{\vspace{3pt}}%
			\endfirsthead
			
			\multicolumn{4}{l}{TABLE~\ref{tab:PGTUnitaryAndPCMv} (continued)}
			\rule{0pt}{12pt}\\
			\noalign{\vspace{1.5pt}}
			\colrule\rule{0pt}{12pt}
			\#&Critical Condition&\makecell[cl]{Massive \\mode}&$b$ sectors\\*
			\colrule
			\noalign{\vspace{3pt}}%
			\endhead
			\noalign{\nobreak\vspace{3pt}}%
			\colrule
			\endfoot
			\noalign{\nobreak\vspace{3pt}}%
			\botrule
			\endlastfoot
			
			\rownumber & \makecell[cl]{$r_1=r_3=r_4=r_5=\lambda =0$}  & $0^-$& \makecell[cl]{$\left\{A\text{}_{\text{v}}^{2},A\text{}_{\text{}}^{0}|\mathfrak{s}\text{}_{\text{l}}^{2},\left(A\text{}_{\text{}}^{0}\&A\text{}_{\text{}}^{0}\right)^\text{N}|\left(A\text{}_{\text{}}^{0}\&\mathfrak{s}\text{}_{\text{l}}^{2}\right)^\text{N}|\left(A\text{}_{\text{}}^{0}\&\mathfrak{a}\text{}_{\text{l}}^{2}\right)^\text{N},\left(A\text{}_{\text{}}^{0}\&A\text{}_{\text{}}^{0}\right)^\text{N}|\left(A\text{}_{\text{}}^{0}\&\mathfrak{a}\text{}_{\text{l}}^{2}\right)^\text{N},A\text{}_{\text{}}^{0},A\text{}_{\text{}}^{0}|\mathfrak{s}\text{}_{\text{l}}^{2}\right\}$} \\
			\rownumber & \makecell[cl]{$r_1=r_3=r_4=r_5=t_1+t_2=\lambda =0$}  &$0^-$& \makecell[cl]{$\left\{A\text{}_{\text{v}}^{2},A\text{}_{\text{}}^{0}|\mathfrak{s}\text{}_{\text{l}}^{2},\left(A\text{}_{\text{}}^{0}\&A\text{}_{\text{}}^{0}\right)^\text{N}|\left(A\text{}_{\text{}}^{0}\&\mathfrak{s}\text{}_{\text{l}}^{2}\right)^\text{N}|\left(A\text{}_{\text{}}^{0}\&\mathfrak{a}\text{}_{\text{l}}^{2}\right)^\text{N},\left(A\text{}_{\text{}}^{\infty}\&A\text{}_{\text{}}^{0}\right)^\text{N}|\left(A\text{}_{\text{}}^{\infty}\&\mathfrak{a}\text{}_{\text{l}}^{2}\right)^\text{N},A\text{}_{\text{}}^{0},A\text{}_{\text{}}^{0}|\mathfrak{s}\text{}_{\text{l}}^{2}\right\}$} \\
			\rownumber & \makecell[cl]{$r_1=r_3=r_4=r_5=t_1+t_3=\lambda =0$}  &$0^-$& \makecell[cl]{$\left\{A\text{}_{\text{v}}^{2},A\text{}_{\text{}}^{0}|\mathfrak{s}\text{}_{\text{l}}^{2},\left(A\text{}_{\text{}}^{\infty}\&A\text{}_{\text{}}^{0}\right)^\text{N}|\left(A\text{}_{\text{}}^{\infty}\&\mathfrak{s}\text{}_{\text{l}}^{2}\right)^\text{N}|\left(A\text{}_{\text{}}^{\infty}\&\mathfrak{a}\text{}_{\text{l}}^{2}\right)^\text{N},\left(A\text{}_{\text{}}^{0}\&A\text{}_{\text{}}^{0}\right)^\text{N}|\left(A\text{}_{\text{}}^{0}\&\mathfrak{a}\text{}_{\text{l}}^{2}\right)^\text{N},A\text{}_{\text{}}^{0},A\text{}_{\text{}}^{0}|\mathfrak{s}\text{}_{\text{l}}^{2}\right\}$} \\
			\rownumber & \makecell[cl]{$r_1=r_3=r_4=r_5=t_1+t_2=t_1+t_3=\lambda =0$}  &$0^-$& \makecell[cl]{$\left\{A\text{}_{\text{v}}^{2},A\text{}_{\text{}}^{0}|\mathfrak{s}\text{}_{\text{l}}^{2},\left(A\text{}_{\text{}}^{\infty}\&A\text{}_{\text{}}^{0}\right)^\text{N}|\left(A\text{}_{\text{}}^{\infty}\&\mathfrak{s}\text{}_{\text{l}}^{2}\right)^\text{N}|\left(A\text{}_{\text{}}^{\infty}\&\mathfrak{a}\text{}_{\text{l}}^{2}\right)^\text{N},\left(A\text{}_{\text{}}^{\infty}\&A\text{}_{\text{}}^{0}\right)^\text{N}|\left(A\text{}_{\text{}}^{\infty}\&\mathfrak{a}\text{}_{\text{l}}^{2}\right)^\text{N},A\text{}_{\text{}}^{0},A\text{}_{\text{}}^{0}|\mathfrak{s}\text{}_{\text{l}}^{2}\right\}$} \\
			\rownumber & \makecell[cl]{$r_1=r_3=r_4=t_1=\lambda =0$}  &$0^-$& \makecell[cl]{$\left\{A\text{}_{\text{v}}^{2},A\text{}_{\text{}}^{0}|\mathfrak{s}\text{}_{\text{l}}^{2},\left(A\text{}_{\text{l}}^{2}\&A\text{}_{\text{l}}^{0}\right)^\text{N}|\left(A\text{}_{\text{l}}^{2}\&\mathfrak{s}\text{}_{\text{l}}^{2}\right)^\text{N}|\left(A\text{}_{\text{l}}^{2}\&\mathfrak{a}\text{}_{\text{l}}^{2}\right)^\text{N},\left(A\text{}_{\text{l}}^{2}\&A\text{}_{\text{l}}^{0}\right)^\text{N}|\left(A\text{}_{\text{l}}^{2}\&\mathfrak{a}\text{}_{\text{l}}^{2}\right)^\text{N},\times,\times\right\}$} \\
			$^{\ast 5}$\rownumber & \makecell[cl]{$r_1=r_3=r_4=r_5=t_1=\lambda =0$}  &$0^-$& $\left\{A\text{}_{\text{v}}^{2},A\text{}_{\text{}}^{0}|\mathfrak{s}\text{}_{\text{l}}^{2},A\text{}_{\text{}}^{0}|\mathfrak{s}\text{}_{\text{l}}^{2}|\mathfrak{a}\text{}_{\text{l}}^{2},A\text{}_{\text{}}^{0}|\mathfrak{a}\text{}_{\text{l}}^{2},\times,\times\right\}$ \\
			$^{\ast 6}$\rownumber & \makecell[cl]{$r_1=r_3=r_4=r_5=t_1=t_3=\lambda =0$}  &$0^-$& $\left\{A\text{}_{\text{v}}^{2},\times,\times,A\text{}_{\text{}}^{0}|\mathfrak{a}\text{}_{\text{l}}^{2},\times,\times\right\}$ \\
			\rownumber & \makecell[cl]{$r_1=\frac{r_3}{2}-r_4=\frac{r_3}{2}+r_5=t_1=t_3=\lambda =0$}  &$0^-$& $\left\{A\text{}_{\text{v}}^{2},\times,\times,\left(A\text{}_{\text{l}}^{2}\&A\text{}_{\text{l}}^{0}\right)^\text{N}|\left(A\text{}_{\text{l}}^{2}\&\mathfrak{a}\text{}_{\text{l}}^{2}\right)^\text{N},\times,A\text{}_{\text{l}}^{2}\right\}$ \\
			\rownumber & \makecell[cl]{$r_1=r_3=r_4=t_1=t_3=\lambda =0$}  &$0^-$& $\left\{A\text{}_{\text{v}}^{2},\times,A\text{}_{\text{l}}^{2},\left(A\text{}_{\text{l}}^{2}\&A\text{}_{\text{l}}^{0}\right)^\text{N}|\left(A\text{}_{\text{l}}^{2}\&\mathfrak{a}\text{}_{\text{l}}^{2}\right)^\text{N},\times,\times\right\}$ \\
			\rownumber & \makecell[cl]{$r_1-r_3=r_4=2 r_1+r_5=t_1=\lambda =0$}  &$0^-$& $\left\{A\text{}_{\text{v}}^{2},A\text{}_{\text{}}^{0}|\mathfrak{s}\text{}_{\text{l}}^{2},\left(A\text{}_{\text{l}}^{2}\&A\text{}_{\text{l}}^{0}\right)^\text{N}|\left(A\text{}_{\text{l}}^{2}\&\mathfrak{s}\text{}_{\text{l}}^{2}\right)^\text{N}|\left(A\text{}_{\text{l}}^{2}\&\mathfrak{a}\text{}_{\text{l}}^{2}\right)^\text{N},A\text{}_{\text{}}^{0}|\mathfrak{a}\text{}_{\text{l}}^{2},A\text{}_{\text{l}}^{2},\times\right\}$ \\
			$^{\ast 7}$\rownumber & \makecell[cl]{$r_1-r_3=r_4=2 r_1+r_5=t_1=t_3=\lambda =0$}  &$0^-$& $\left\{A\text{}_{\text{v}}^{2},\times,A\text{}_{\text{l}}^{2},A\text{}_{\text{}}^{0}|\mathfrak{a}\text{}_{\text{l}}^{2},A\text{}_{\text{l}}^{2},\times\right\}$ \\
			$^{\ast 8}$\rownumber & \makecell[cl]{$r_1=2 r_3-r_4=2 r_3+r_5=t_1=t_3=\lambda =0$}  &$0^-$& $\left\{A\text{}_{\text{v}}^{2},A\text{}_{\text{l}}^{2},\times,A\text{}_{\text{}}^{0}|\mathfrak{a}\text{}_{\text{l}}^{2},\times,\times\right\}$ \\
			\rownumber & \makecell[cl]{$r_1=r_3=r_4=r_5=t_3=\lambda =0$}  &$0^-$& $\left\{A\text{}_{\text{v}}^{2},\times,A\text{}_{\text{}}^{0}|\mathfrak{s}\text{}_{\text{l}}^{2}|\mathfrak{a}\text{}_{\text{l}}^{2},\left(A\text{}_{\text{}}^{0}\&A\text{}_{\text{}}^{0}\right)^\text{N}|\left(A\text{}_{\text{}}^{0}\&\mathfrak{a}\text{}_{\text{l}}^{2}\right)^\text{N},A\text{}_{\text{}}^{0},A\text{}_{\text{}}^{0}|\mathfrak{s}\text{}_{\text{l}}^{2}\right\}$ \\
			\rownumber & \makecell[cl]{$r_1=r_3=r_4=r_5=t_1+t_2=t_3=\lambda =0$}  &$0^-$& $\left\{A\text{}_{\text{v}}^{2},\times,A\text{}_{\text{}}^{0}|\mathfrak{s}\text{}_{\text{l}}^{2}|\mathfrak{a}\text{}_{\text{l}}^{2},\left(A\text{}_{\text{}}^{\infty}\&A\text{}_{\text{}}^{0}\right)^\text{N}|\left(A\text{}_{\text{}}^{\infty}\&\mathfrak{a}\text{}_{\text{l}}^{2}\right)^\text{N},A\text{}_{\text{}}^{0},A\text{}_{\text{}}^{0}|\mathfrak{s}\text{}_{\text{l}}^{2}\right\}$ \\
			\rownumber & \makecell[cl]{$r_1=2 r_3-r_4=t_1=t_3=\lambda =0$}  &$0^-$& $\left\{A\text{}_{\text{v}}^{2},A\text{}_{\text{l}}^{2},A\text{}_{\text{l}}^{2},\left(A\text{}_{\text{l}}^{2}\&A\text{}_{\text{l}}^{0}\right)^\text{N}|\left(A\text{}_{\text{l}}^{2}\&\mathfrak{a}\text{}_{\text{l}}^{2}\right)^\text{N},\times,\times\right\}$ \\
			$^{\ast 9}$\rownumber & \makecell[cl]{$r_1=\frac{r_3}{2}-r_4=2 r_3+r_5=t_1=t_3=\lambda =0$}  &$0^-$& $\left\{A\text{}_{\text{v}}^{2},\times,A\text{}_{\text{l}}^{2},A\text{}_{\text{}}^{0}|\mathfrak{a}\text{}_{\text{l}}^{2},\times,A\text{}_{\text{l}}^{2}\right\}$ \\
			$^{\ast 10}$\rownumber & \makecell[cl]{$2 r_1-2 r_3+r_4=2 r_3+r_5=t_1=t_3=\lambda =0$}  &$0^-$& $\left\{A\text{}_{\text{v}}^{2},A\text{}_{\text{l}}^{2},A\text{}_{\text{l}}^{2},A\text{}_{\text{}}^{0}|\mathfrak{a}\text{}_{\text{l}}^{2},A\text{}_{\text{l}}^{2},\times\right\}$ \\
			\rownumber & \makecell[cl]{$r_1=\frac{r_3}{2}-r_4=2 r_3+r_5=t_1=\lambda =0$}  &$0^-$& $\left\{A\text{}_{\text{v}}^{2},A\text{}_{\text{}}^{0}|\mathfrak{s}\text{}_{\text{l}}^{2},\left(A\text{}_{\text{l}}^{2}\&A\text{}_{\text{l}}^{0}\right)^\text{N}|\left(A\text{}_{\text{l}}^{2}\&\mathfrak{s}\text{}_{\text{l}}^{2}\right)^\text{N}|\left(A\text{}_{\text{l}}^{2}\&\mathfrak{a}\text{}_{\text{l}}^{2}\right)^\text{N},A\text{}_{\text{}}^{0}|\mathfrak{a}\text{}_{\text{l}}^{2},\times,A\text{}_{\text{l}}^{2}\right\}$ \\
			\rownumber & \makecell[cl]{$r_1=2 r_3-r_4=2 r_3+r_5=t_3=\lambda =0$}  &$0^-$& $\left\{A\text{}_{\text{v}}^{2},A\text{}_{\text{l}}^{2},A\text{}_{\text{}}^{0}|\mathfrak{s}\text{}_{\text{l}}^{2}|\mathfrak{a}\text{}_{\text{l}}^{2},\left(A\text{}_{\text{}}^{0}\&A\text{}_{\text{}}^{0}\right)^\text{N}|\left(A\text{}_{\text{}}^{0}\&\mathfrak{a}\text{}_{\text{l}}^{2}\right)^\text{N},A\text{}_{\text{}}^{0},A\text{}_{\text{}}^{0}|\mathfrak{s}\text{}_{\text{l}}^{2}\right\}$ \\
			\rownumber & \makecell[cl]{$r_1=2 r_3-r_4=2 r_3+r_5=t_1+t_2=t_3=\lambda =0$}  &$0^-$& $\left\{A\text{}_{\text{v}}^{2},A\text{}_{\text{l}}^{2},A\text{}_{\text{}}^{0}|\mathfrak{s}\text{}_{\text{l}}^{2}|\mathfrak{a}\text{}_{\text{l}}^{2},\left(A\text{}_{\text{}}^{\infty}\&A\text{}_{\text{}}^{0}\right)^\text{N}|\left(A\text{}_{\text{}}^{\infty}\&\mathfrak{a}\text{}_{\text{l}}^{2}\right)^\text{N},A\text{}_{\text{}}^{0},A\text{}_{\text{}}^{0}|\mathfrak{s}\text{}_{\text{l}}^{2}\right\}$ \\
			\rownumber & \makecell[cl]{$r_1=r_4+r_5=t_1=t_3=\lambda =0$}  &$0^-$& $\left\{A\text{}_{\text{v}}^{2},A\text{}_{\text{l}}^{2},\times,\left(A\text{}_{\text{l}}^{2}\&A\text{}_{\text{l}}^{0}\right)^\text{N}|\left(A\text{}_{\text{l}}^{2}\&\mathfrak{a}\text{}_{\text{l}}^{2}\right)^\text{N},\times,A\text{}_{\text{l}}^{2}\right\}$ \\
			\rownumber & \makecell[cl]{$r_1=\frac{r_3}{2}-r_4=\frac{r_3}{2}+r_5=t_1=\lambda =0$}  &$0^-$& $\left\{A\text{}_{\text{v}}^{2},A\text{}_{\text{}}^{0}|\mathfrak{s}\text{}_{\text{l}}^{2},A\text{}_{\text{}}^{0}|\mathfrak{s}\text{}_{\text{l}}^{2}|\mathfrak{a}\text{}_{\text{l}}^{2},\left(A\text{}_{\text{l}}^{2}\&A\text{}_{\text{l}}^{0}\right)^\text{N}|\left(A\text{}_{\text{l}}^{2}\&\mathfrak{a}\text{}_{\text{l}}^{2}\right)^\text{N},\times,A\text{}_{\text{l}}^{2}\right\}$ \\
			
			\rownumber & \makecell[cl]{$r_1,r_3,r_4,t_1+t_2,\lambda =0$} &$0^-,1^-$ & $\left\{A\text{}_{\text{v}}^{2},A\text{}_{\text{}}^{0}|\mathfrak{s}\text{}_{\text{l}}^{2},\left(A\text{}_{\text{v}}^{2}\&A\text{}_{\text{v}}^{0}\right)^\text{N}|\left(A\text{}_{\text{v}}^{2}\&\mathfrak{s}\text{}_{\text{vl}}^{2}\right)^\text{N}|\left(A\text{}_{\text{v}}^{2}\&\mathfrak{a}\text{}_{\text{vl}}^{2}\right)^\text{N},\left(A\text{}_{\text{}}^{\infty}\&A\text{}_{\text{}}^{-2}\right)^\text{N}|\left(A\text{}_{\text{}}^{\infty}\&\mathfrak{a}\text{}_{\text{l}}^{0}\right)^\text{N},A\text{}_{\text{}}^{0},A\text{}_{\text{}}^{0}|\mathfrak{s}\text{}_{\text{l}}^{2}\right\}$ \\
			\rownumber & \makecell[cl]{$r_1,r_3,r_4,t_1+t_3,\lambda =0$} &$0^-,1^+$ & $\left\{A\text{}_{\text{v}}^{2},A\text{}_{\text{}}^{0}|\mathfrak{s}\text{}_{\text{l}}^{2},\left(A\text{}_{\text{}}^{\infty}\&A\text{}_{\text{}}^{-2}\right)^\text{N}|\left(A\text{}_{\text{}}^{\infty}\&\mathfrak{s}\text{}_{\text{l}}^{0}\right)^\text{N}|\left(A\text{}_{\text{}}^{\infty}\&\mathfrak{a}\text{}_{\text{l}}^{0}\right)^\text{N},\left(A\text{}_{\text{v}}^{2}\&A\text{}_{\text{v}}^{0}\right)^\text{N}|\left(A\text{}_{\text{v}}^{2}\&\mathfrak{a}\text{}_{\text{vl}}^{2}\right)^\text{N},A\text{}_{\text{}}^{0},A\text{}_{\text{}}^{0}|\mathfrak{s}\text{}_{\text{l}}^{2}\right\}$ \\
			\rownumber & \makecell[cl]{$r_2,r_1-r_3,r_4,t_1+t_2,\lambda =0$} &$1^-,2^-$ & $\left\{A\text{}_{\text{}}^{0},A\text{}_{\text{}}^{0}|\mathfrak{s}\text{}_{\text{l}}^{2},\left(A\text{}_{\text{v}}^{2}\&A\text{}_{\text{v}}^{0}\right)^\text{N}|\left(A\text{}_{\text{v}}^{2}\&\mathfrak{s}\text{}_{\text{vl}}^{2}\right)^\text{N}|\left(A\text{}_{\text{v}}^{2}\&\mathfrak{a}\text{}_{\text{vl}}^{2}\right)^\text{N},\left(A\text{}_{\text{}}^{\infty}\&A\text{}_{\text{}}^{-2}\right)^\text{N}|\left(A\text{}_{\text{}}^{\infty}\&\mathfrak{a}\text{}_{\text{l}}^{0}\right)^\text{N},A\text{}_{\text{v}}^{2},A\text{}_{\text{}}^{0}|\mathfrak{s}\text{}_{\text{l}}^{2}\right\}$ \\
			\rownumber & \makecell[cl]{$r_2,r_1-r_3,r_4,t_1+t_3,\lambda =0$} &$1^+,2^-$ & $\left\{A\text{}_{\text{}}^{0},A\text{}_{\text{}}^{0}|\mathfrak{s}\text{}_{\text{l}}^{2},\left(A\text{}_{\text{}}^{\infty}\&A\text{}_{\text{}}^{-2}\right)^\text{N}|\left(A\text{}_{\text{}}^{\infty}\&\mathfrak{s}\text{}_{\text{l}}^{0}\right)^\text{N}|\left(A\text{}_{\text{}}^{\infty}\&\mathfrak{a}\text{}_{\text{l}}^{0}\right)^\text{N},\left(A\text{}_{\text{v}}^{2}\&A\text{}_{\text{v}}^{0}\right)^\text{N}|\left(A\text{}_{\text{v}}^{2}\&\mathfrak{a}\text{}_{\text{vl}}^{2}\right)^\text{N},A\text{}_{\text{v}}^{2},A\text{}_{\text{}}^{0}|\mathfrak{s}\text{}_{\text{l}}^{2}\right\}$ \\
			\rownumber & \makecell[cl]{$r_1-r_3,r_4,2 r_1+r_5,t_1+t_3,\lambda =0$} &$0^-,2^-$ & $\left\{A\text{}_{\text{v}}^{2},A\text{}_{\text{}}^{0}|\mathfrak{s}\text{}_{\text{l}}^{2},\left(A\text{}_{\text{}}^{\infty}\&A\text{}_{\text{}}^{-2}\right)^\text{N}|\left(A\text{}_{\text{}}^{\infty}\&\mathfrak{s}\text{}_{\text{l}}^{0}\right)^\text{N}|\left(A\text{}_{\text{}}^{\infty}\&\mathfrak{a}\text{}_{\text{l}}^{0}\right)^\text{N},\left(A\text{}_{\text{}}^{0}\&A\text{}_{\text{}}^{0}\right)^\text{N}|\left(A\text{}_{\text{}}^{0}\&\mathfrak{a}\text{}_{\text{l}}^{2}\right)^\text{N},A\text{}_{\text{v}}^{2},A\text{}_{\text{}}^{0}|\mathfrak{s}\text{}_{\text{l}}^{2}\right\}$ \\
			\rownumber & \makecell[cl]{$r_1,r_2,r_3,r_4,t_1+t_2,\lambda =0$} &$1^-$ & $\left\{A\text{}_{\text{}}^{0},A\text{}_{\text{}}^{0}|\mathfrak{s}\text{}_{\text{l}}^{2},\left(A\text{}_{\text{v}}^{2}\&A\text{}_{\text{v}}^{0}\right)^\text{N}|\left(A\text{}_{\text{v}}^{2}\&\mathfrak{s}\text{}_{\text{vl}}^{2}\right)^\text{N}|\left(A\text{}_{\text{v}}^{2}\&\mathfrak{a}\text{}_{\text{vl}}^{2}\right)^\text{N},\left(A\text{}_{\text{}}^{\infty}\&A\text{}_{\text{}}^{-2}\right)^\text{N}|\left(A\text{}_{\text{}}^{\infty}\&\mathfrak{a}\text{}_{\text{l}}^{0}\right)^\text{N},A\text{}_{\text{}}^{0},A\text{}_{\text{}}^{0}|\mathfrak{s}\text{}_{\text{l}}^{2}\right\}$ \\
			\rownumber & \makecell[cl]{$r_1,r_2,r_3,r_4,t_1+t_3,\lambda =0$} &$1^+$ & $\left\{A\text{}_{\text{}}^{0},A\text{}_{\text{}}^{0}|\mathfrak{s}\text{}_{\text{l}}^{2},\left(A\text{}_{\text{}}^{\infty}\&A\text{}_{\text{}}^{-2}\right)^\text{N}|\left(A\text{}_{\text{}}^{\infty}\&\mathfrak{s}\text{}_{\text{l}}^{0}\right)^\text{N}|\left(A\text{}_{\text{}}^{\infty}\&\mathfrak{a}\text{}_{\text{l}}^{0}\right)^\text{N},\left(A\text{}_{\text{v}}^{2}\&A\text{}_{\text{v}}^{0}\right)^\text{N}|\left(A\text{}_{\text{v}}^{2}\&\mathfrak{a}\text{}_{\text{vl}}^{2}\right)^\text{N},A\text{}_{\text{}}^{0},A\text{}_{\text{}}^{0}|\mathfrak{s}\text{}_{\text{l}}^{2}\right\}$ \\
			\rownumber & \makecell[cl]{$r_1,r_3,r_4,t_1+t_2,t_1+t_3,\lambda =0$} &$0^-$ & $\left\{A\text{}_{\text{v}}^{2},A\text{}_{\text{}}^{0}|\mathfrak{s}\text{}_{\text{l}}^{2},\left(A\text{}_{\text{}}^{\infty}\&A\text{}_{\text{}}^{-2}\right)^\text{N}|\left(A\text{}_{\text{}}^{\infty}\&\mathfrak{s}\text{}_{\text{l}}^{0}\right)^\text{N}|\left(A\text{}_{\text{}}^{\infty}\&\mathfrak{a}\text{}_{\text{l}}^{0}\right)^\text{N},\left(A\text{}_{\text{}}^{\infty}\&A\text{}_{\text{}}^{-2}\right)^\text{N}|\left(A\text{}_{\text{}}^{\infty}\&\mathfrak{a}\text{}_{\text{l}}^{0}\right)^\text{N},A\text{}_{\text{}}^{0},A\text{}_{\text{}}^{0}|\mathfrak{s}\text{}_{\text{l}}^{2}\right\}$ \\
			\rownumber & \makecell[cl]{$r_2,r_1-r_3,r_4,r_1+r_5,t_1+t_2,\lambda =0$} &$2^-$ & $\left\{A\text{}_{\text{}}^{0},A\text{}_{\text{}}^{0}|\mathfrak{s}\text{}_{\text{l}}^{2},\left(A\text{}_{\text{}}^{0}\&A\text{}_{\text{}}^{0}\right)^\text{N}|\left(A\text{}_{\text{}}^{0}\&\mathfrak{s}\text{}_{\text{l}}^{2}\right)^\text{N}|\left(A\text{}_{\text{}}^{0}\&\mathfrak{a}\text{}_{\text{l}}^{2}\right)^\text{N},\left(A\text{}_{\text{}}^{\infty}\&A\text{}_{\text{}}^{-2}\right)^\text{N}|\left(A\text{}_{\text{}}^{\infty}\&\mathfrak{a}\text{}_{\text{l}}^{0}\right)^\text{N},A\text{}_{\text{v}}^{2},A\text{}_{\text{}}^{0}|\mathfrak{s}\text{}_{\text{l}}^{2}\right\}$ \\
			\rownumber & \makecell[cl]{$r_2,r_1-r_3,r_4,2 r_1+r_5,t_1+t_3,\lambda =0$} &$2^-$ & $\left\{A\text{}_{\text{}}^{0},A\text{}_{\text{}}^{0}|\mathfrak{s}\text{}_{\text{l}}^{2},\left(A\text{}_{\text{}}^{\infty}\&A\text{}_{\text{}}^{-2}\right)^\text{N}|\left(A\text{}_{\text{}}^{\infty}\&\mathfrak{s}\text{}_{\text{l}}^{0}\right)^\text{N}|\left(A\text{}_{\text{}}^{\infty}\&\mathfrak{a}\text{}_{\text{l}}^{0}\right)^\text{N},\left(A\text{}_{\text{}}^{0}\&A\text{}_{\text{}}^{0}\right)^\text{N}|\left(A\text{}_{\text{}}^{0}\&\mathfrak{a}\text{}_{\text{l}}^{2}\right)^\text{N},A\text{}_{\text{v}}^{2},A\text{}_{\text{}}^{0}|\mathfrak{s}\text{}_{\text{l}}^{2}\right\}$ \\
			\rownumber & \makecell[cl]{$r_2,r_1-r_3,r_4,t_1+t_2,t_1+t_3,\lambda =0$} &$2^-$ & $\left\{A\text{}_{\text{}}^{0},A\text{}_{\text{}}^{0}|\mathfrak{s}\text{}_{\text{l}}^{2},\left(A\text{}_{\text{}}^{\infty}\&A\text{}_{\text{}}^{-2}\right)^\text{N}|\left(A\text{}_{\text{}}^{\infty}\&\mathfrak{s}\text{}_{\text{l}}^{0}\right)^\text{N}|\left(A\text{}_{\text{}}^{\infty}\&\mathfrak{a}\text{}_{\text{l}}^{0}\right)^\text{N},\left(A\text{}_{\text{}}^{\infty}\&A\text{}_{\text{}}^{-2}\right)^\text{N}|\left(A\text{}_{\text{}}^{\infty}\&\mathfrak{a}\text{}_{\text{l}}^{0}\right)^\text{N},A\text{}_{\text{v}}^{2},A\text{}_{\text{}}^{0}|\mathfrak{s}\text{}_{\text{l}}^{2}\right\}$ \\
			\rownumber & \makecell[cl]{$r_2,r_1-r_3,r_4,r_1+r_5,t_1+t_2,t_1+t_3,\lambda =0$} &$2^-$ & $\left\{A\text{}_{\text{}}^{0},A\text{}_{\text{}}^{0}|\mathfrak{s}\text{}_{\text{l}}^{2},\left(A\text{}_{\text{}}^{\infty}\&A\text{}_{\text{}}^{0}\right)^\text{N}|\left(A\text{}_{\text{}}^{\infty}\&\mathfrak{s}\text{}_{\text{l}}^{2}\right)^\text{N}|\left(A\text{}_{\text{}}^{\infty}\&\mathfrak{a}\text{}_{\text{l}}^{2}\right)^\text{N},\left(A\text{}_{\text{}}^{\infty}\&A\text{}_{\text{}}^{-2}\right)^\text{N}|\left(A\text{}_{\text{}}^{\infty}\&\mathfrak{a}\text{}_{\text{l}}^{0}\right)^\text{N},A\text{}_{\text{v}}^{2},A\text{}_{\text{}}^{0}|\mathfrak{s}\text{}_{\text{l}}^{2}\right\}$ \\
			\rownumber & \makecell[cl]{$r_2,r_1-r_3,r_4,2 r_1+r_5,t_1+t_2,t_1+t_3,\lambda =0$} &$2^-$ & $\left\{A\text{}_{\text{}}^{0},A\text{}_{\text{}}^{0}|\mathfrak{s}\text{}_{\text{l}}^{2},\left(A\text{}_{\text{}}^{\infty}\&A\text{}_{\text{}}^{-2}\right)^\text{N}|\left(A\text{}_{\text{}}^{\infty}\&\mathfrak{s}\text{}_{\text{l}}^{0}\right)^\text{N}|\left(A\text{}_{\text{}}^{\infty}\&\mathfrak{a}\text{}_{\text{l}}^{0}\right)^\text{N},\left(A\text{}_{\text{}}^{\infty}\&A\text{}_{\text{}}^{0}\right)^\text{N}|\left(A\text{}_{\text{}}^{\infty}\&\mathfrak{a}\text{}_{\text{l}}^{2}\right)^\text{N},A\text{}_{\text{v}}^{2},A\text{}_{\text{}}^{0}|\mathfrak{s}\text{}_{\text{l}}^{2}\right\}$ \\
			\rownumber & \makecell[cl]{$r_2,r_1-r_3,r_4,r_1+r_5,t_1+t_2,t_3,\lambda =0$} &$2^-$ & $\left\{A\text{}_{\text{}}^{0},\times,A\text{}_{\text{}}^{0}|\mathfrak{s}\text{}_{\text{l}}^{2}|\mathfrak{a}\text{}_{\text{l}}^{2},\left(A\text{}_{\text{}}^{\infty}\&A\text{}_{\text{}}^{-2}\right)^\text{N}|\left(A\text{}_{\text{}}^{\infty}\&\mathfrak{a}\text{}_{\text{l}}^{0}\right)^\text{N},A\text{}_{\text{v}}^{2},A\text{}_{\text{}}^{0}|\mathfrak{s}\text{}_{\text{l}}^{2}\right\}$ \\
			\rownumber & \makecell[cl]{$r_2,r_1-r_3,r_4,2 r_1+r_5,t_2,t_1+t_3,\lambda =0$} &$2^-$ & $\left\{\times,A\text{}_{\text{}}^{0}|\mathfrak{s}\text{}_{\text{l}}^{2},\left(A\text{}_{\text{}}^{\infty}\&A\text{}_{\text{}}^{-2}\right)^\text{N}|\left(A\text{}_{\text{}}^{\infty}\&\mathfrak{s}\text{}_{\text{l}}^{0}\right)^\text{N}|\left(A\text{}_{\text{}}^{\infty}\&\mathfrak{a}\text{}_{\text{l}}^{0}\right)^\text{N},A\text{}_{\text{}}^{0}|\mathfrak{a}\text{}_{\text{l}}^{2},A\text{}_{\text{v}}^{2},A\text{}_{\text{}}^{0}|\mathfrak{s}\text{}_{\text{l}}^{2}\right\}$ \\
			\rownumber & \makecell[cl]{$r_1-r_3,r_4,2 r_1+r_5,t_2,t_1+t_3,\lambda =0$} &$2^-$ & $\left\{A\text{}_{\text{l}}^{2},A\text{}_{\text{}}^{0}|\mathfrak{s}\text{}_{\text{l}}^{2},\left(A\text{}_{\text{}}^{\infty}\&A\text{}_{\text{}}^{-2}\right)^\text{N}|\left(A\text{}_{\text{}}^{\infty}\&\mathfrak{s}\text{}_{\text{l}}^{0}\right)^\text{N}|\left(A\text{}_{\text{}}^{\infty}\&\mathfrak{a}\text{}_{\text{l}}^{0}\right)^\text{N},A\text{}_{\text{}}^{0}|\mathfrak{a}\text{}_{\text{l}}^{2},A\text{}_{\text{v}}^{2},A\text{}_{\text{}}^{0}|\mathfrak{s}\text{}_{\text{l}}^{2}\right\}$ \\
			\rownumber & \makecell[cl]{$r_2,2 r_1-2 r_3+r_4,r_1-2 r_3-r_5,t_1+t_2,t_3,\lambda =0$} &$2^-$ & $\left\{A\text{}_{\text{}}^{0},A\text{}_{\text{l}}^{2},A\text{}_{\text{}}^{0}|\mathfrak{s}\text{}_{\text{l}}^{2}|\mathfrak{a}\text{}_{\text{l}}^{2},\left(A\text{}_{\text{}}^{\infty}\&A\text{}_{\text{}}^{-2}\right)^\text{N}|\left(A\text{}_{\text{}}^{\infty}\&\mathfrak{a}\text{}_{\text{l}}^{0}\right)^\text{N},A\text{}_{\text{v}}^{2},A\text{}_{\text{}}^{0}|\mathfrak{s}\text{}_{\text{l}}^{2}\right\}$ \\
	\end{longtable*}
	\egroup
\FloatBarrier
	\bibliographystyle{apsrev4-2}
	\bibliography{GaugeGravity_abbr_using}

\end{document}
%